%% file: main.tex
\documentclass[lettersize,journal]{IEEEtran}
\input{preamble}

\def\BibTeX{{\rm B\kern-.05em{\sc i\kern-.025em b}\kern-.08em
    T\kern-.1667em\lower.7ex\hbox{E}\kern-.125emX}}

\begin{document}
%
%
\title{Performance Analysis and Evaluation of Post Quantum Secure Blockchained Federated
Learning
}

\author{Dev Gurung, Shiva Raj Pokhrel, Gang Li
}

\maketitle

\input{abstract}


\input{body}



\clearpage

\section{APPENDIX}
\subsection{Benefits of a PQC-PQC approach}
\begin{prop}
The PQC-PQC approach is more secure than PQC alone or classical-PQC combination i.e.
\begin{equation}
    S(PQC-PQC) > S(PQC) \text{ or } S(classical-PQC)
\end{equation}
\end{prop}

The main idea of using the PQC-PQC combination of two PQC algorithms together in a cryptographic system is to leverage the strengths and complement the drawbacks of both PQCs and create a more secure and future-proof system.
Furthermore, our proposed method is not just a PQC-PQC hybrid implementation.
Our implementation is specifically the XMMS-dilithium or XMMS-Falcon hybrid implementation. 
That means, the implementation is a hybrid approach specifically for the stateful hash-based signature scheme, XMSS with a stateless lattice-based signature scheme Dilithium or Falcon.

\subsection{Goods or Bads of XMSS}
\begin{enumerate}
    \item Stateful: XMSS is a stateful signature scheme. This means that the number of keys is limited and that a single key cannot be used more than once.
    \item Synchronization Issues: In blockchains such as QRL \cite{QuantumResistantLedger2016} that implements XMSS, there is a typical issue. Once the user runs out of keys to sign the transitions, a new XMSS tree needs to be generated. 
    If the new account generation is not timely and the old account balance is not transferred with the last key, then the balance will be lost. This could be a major issue that needs to be addressed.
    \item Minimum Security Assumptions: Hash-based signature schemes are based on minimum security assumptions i.e. they are just based on hash-based functions. Thus, it does not require major hardware improvement or cryptanalysis to understand that it is scalable, suitable, and quantum secure.
    \item Fast Performance: The performance of the XMSS signature scheme is very good in comparison to other PQC schemes. With a small height as recommended by our proposed method, XMSS schemes will be overall best in performance.   
\end{enumerate}

\subsection{Goods and Bads of Dilithium or Falcon}
\begin{enumerate}
    \item Efficient: Both Dilithium and Falcon are promising signature schemes with very good key generation and signing performance. 
    Their performance is comparable to that of classical RSA-type cryptography. However, they provide quantum security and RSA doesn't. 
    \item Signature \& Public key sizes: Both schemes have significantly larger public and signature key sizes than traditional signature schemes such as RSA, etc.
\end{enumerate}

\subsection{Security Analysis for XMMS-Dilithium Signature Scheme}
\subsubsection{Dilithium Security}
The most common security concept for digital signatures is UF-CMA (Unforgeability under Choice Message Attacks), where the attacker has access to the signing oracle and can sign messages of their choice using a public key.
The attacker's goal is to produce a valid signature for a new message. 
A stronger security requirement, SUF-CMA (Strong Unforgeability Under Chosen Message Attacks), allows the attacker to win by creating a different signature for an already known message.
Dilithium has been proven SUF-CMA secure in the (classical) random oracle model, based on the difficulty of standard mathematical problems like MLWE and MSIS.

Dilithium is based on the following hardness assumptions \cite{ducasCRYSTALSDilithiumLatticeBasedDigital2018}.
\begin{enumerate}
    \item \textbf{MLWE Problem.} With $m$ and $k$ as integers, let $D: \mathbb{R}q \rightarrow [0, 1]$ be a probability distribution. The advantage of the $A$ algorithm in solving the problem MLWE$_{m,k,D}$ over $Z_q$ ring is defined as:
    \begin{align*}
        \mathrm{Adv}^{MLWE}_{m,k,D} 
        &:= 
         \vert \mathrm{Prob}
        [b=1 | A \gets \mathbb{Z}_q^{m \times k}, t \gets \mathbb{Z}_q^m, b \gets A(A,t)
        ] 
         \\
        & \qquad - 
        \mathrm{Prob}
        [b=1 | A \gets \mathbb{Z}_q^{m\times k};  \\
        & s_1 \gets D^k; s_2 \gets D^m; b \gets A(A,As1+s2)] \vert. 
    \end{align*}

    \item \textbf{MSIS Problem.} For alorithm $A$, $\mathrm{Adv}^{MSIS}_{m,k,\tau}$ to solve MSIS$_{m,k,\tau}$ over $R_q$ ring can be stated as follows: 
    
    \begin{align*}
    \mathrm{Adv}^{MSIS}_{m,k,\tau}(A) 
    & := \mathrm{Prob}
    [0 < ||y||_{\infty} 
    \leq \tau \wedge [I \vert A] \cdot y  \\
    & = 0 | A \gets \mathbb{Z}_q^{m\times k};y \gets A(\mathbb{A})]
    \end{align*}

    \item \textbf{SelfTargetMSIS Problem.} Lets assume $H : \{0, 1\}^* \rightarrow B_{60}$, a cryptographic hash function then we can associate the advantage function $\mathrm{Adv}^{SelfTargetMSIS}_{H,m,k,\tau}(A)$ to an algorithm $A$ in order to solve the Self-Targeting MSIS$_{m,k,\tau}$ problem over ring $Z_q$.

    \begin{align*}
    \mathrm{Adv}^{SelfTargetMSIS}_{H,m,k,\tau}(A) 
    & := \mathrm{Prob}
    \bigl[0 \leq ||y||_{\infty} \leq \tau \wedge \text{H} \\
    & ([I\vert A]\cdot y\Vert M) = c | A \gets \mathbb{Z}_q^{m\times k};\\
    &\left(y:= \begin{bmatrix} r \\ c \end{bmatrix},M\right) \gets A^{\vert H(\cdot)\rangle}(A)\bigr].
    \end{align*}
\end{enumerate}

Thus, from \cite{kiltzConcreteTreatmentFiatShamir2017a},
QROM security of Dilithium can be as \cite{ducasCRYSTALSDilithiumLatticeBasedDigital2018}:
\begin{align*}
    \text{Adv}^{\text{SUF-CMA}}_{\text{Dilithium}}(A) 
    & \leq \text{Adv}^{\text{MLWE}}_{k, l ,D}(B) + \\
    & \text{Adv}^{\text{SelfTargetMSIS}}_{H,k,l+1,\lambda}(C) + \\
     & \text{Adv}^{\text{MSIS}}_{k, l ,\lambda'}(D) + 2^{-256}
\end{align*}
where uniform distribution over $S_\eta$ is $D$,
$$\lambda = \max\{\tau_1 - \beta, 2\tau_2 + 1 + 
2^{d-1}\cdot 60\} \leq 4\tau_2,$$
$$\lambda' = \max\{2(\tau' - \beta), 4\tau_2 + 2\} 
\leq 4\tau_2 + 2.$$

\subsection{Hybrid Scheme Analysis} 
With most PQC schemes, their implementation is bound to affect
the overall performance of the system.
With their larger key and signature sizes along with slower
key generation time and verification time, it is important to
understand and experiment with how they work in different scenarios.

\begin{prop}
Dilithium and XMSS are quantum T-secure. Thus, the proposed hybrid signature is also quantum secure.
\end{prop}

The proposed hybrid scheme has a triple $$ \left (KeyGen (D, X), Sign (X), V (V (X, D) V (X)))\right.$$

$KeyGen(D,X)$ refers to key generation by Dilithium and XMSS, $sign(X)$ refers to signing transactions using XMSS, and verification involves verification of XMSS signatures of transactions and
verification of the dilithium signature of the XMSS public key.
$V(X,D)$ refers to the verification of the Dilithium signature of the XMSS public key, and
$V(X)$ refers to the verification of the XMSS signatures.
So, we can write
$$\forall (dKeys, xKeys) \leftarrow KeyGen (1{\lambda_x, \lambda_d})$$
$$\forall (m \in M): V(V_d(dPk, X_pk, d_\delta), V_x (tx, X_{pk}, x_\delta)) = 1$$ 
where $\lambda_d, \lambda_x$ are security parameters for Dilithium and XMSS, respectively.



\subsection{Convergence Analysis}
Theoretical Analysis of convergence behavior of the role selection mechanism algorithm.
\textbf{Assumption:}[Gradient Approximation Error]
In each round of communication $r$, the role selection algorithm selects the set of $S_d$ online devices in a way such that the aggregated gradients of workers are a good approximation with error $\epsilon$.

\begin{prop}
Consensus Convergence Time in BFL with the Proposed Miner Selection.
\end{prop}
\begin{proof}
In proof of work, we calculate a hash value of the block with a random nonce value that is less than a
threshold value pre-defined.
$$Hash(nonce || Hash(block)) < threshold$$ which takes around 10 minutes for Bitcoin.
This time is dependent on the difficulty level, which refers to the number of zeros at the front of the hash value.
Let the difficulty value be $d$ which is associated with several values such that $$ d \propto numberOfZeros$$
In the stake proof, time is not dependent on computational power.

In BFL with proposed role selection, PoS consensus is followed by final block selection of different
miners comparing two VRF outputs.
Therefore, consensus convergence occurs in the time it takes to compute the central VRF output and the comparison
between the initial distributed VRF output value of each device i.e.,
$$T_{c} \sim Time(T_{vc} + T_{cvrf})$$
where, $T_c$ is the time for consensus convergence, $T_{vc}$ is the comparison of the VRF output and $T_{cvrf}$ is the time taken to compute the VRF output for
central VRF.
To put things in perspective, $T_{c}$ for Bitcoin is around 10 min, while Ethereum takes only 14 seconds. 
\end{proof}

\begin{prop}
VRF added a verification requirement unbreakable by Quantum Attacks.
\end{prop}
\begin{proof}
WOTS+ signing and verification introduced to ECVRF add quantum security.
As shown in Definition \ref{definition:mvrf}, for verification step 2, WOTS+ verification is required.
Even if ecVRF in itself is broken by Quantum Attacks, the WOTS+ signature cannot be forged.
Also, in each round $i$, 
$$sk_i, pk_i \leftarrow wotsKeyGen(randomSeed) $$
So, every time the signing and verification with a specific key are only for that particular communication round.
\end{proof}

\noindent \textbf{Final Miner Selection Using VRF.}\\
Our role selection mechanism already selects miners
mostly with the highest stake. 
Thus, we can simply select any miner to mine the
final block.
However, to make it fair, we use VRF to randomly select
the miner.
For that purpose, we compare our initial VRF output that
we calculated in the beginning to select a role
for the device with another VRF output that is calculated
at the time of mining.
The only difference is that this will be computed once
for everyone.
The miner with the initial VRF value closest to this
new VRF output will be selected as the winning miner.

\begin{prop}
Using VRF to select candidate blocks leverages the performance of BFL based on PoS preventing less forking.
\end{prop}
\begin{proof}
Suppose that we have $m$ the number of miners. 
Then, if we want to decide which miner's block to be appended following Proof of Work, then the time $t$ required to achieve consensus is proportional to computational power (or hash power) $c$, that is,
$$t \propto c$$
To find the nonce in PoW, 
$$ t \propto hashingFunction() \text{ . } difficulty$$
In Proof of Stake, 
$$t \propto coinAge$$ or $$t \propto randomFunctionSpeed()$$
In our proposed method, we use VRF to select candidate blocks. 
Thus, 
$$t \propto VRFfunctionSpeed()$$

Here, VRFfunctionSpeed() is the only time it takes to select a candidate block.
In traditional PoS, the time depends on many other factors, such as randomness, age of the coin, etc. For the forking event to occur, two or more miners blocks are appended by different groups of clients.
Let us suppose $nM$ number of miners and $nD$ devices.
The probability of a forking event occurring will be $$ P(forking) \propto \frac{nfM}{nM} \frac{nfD}{nD}$$ where, 
$nfM$ is the number of miners block appended by different devices and
$nfD$ refers to the number of devices that appended different miners' blocks causing forking.
With the proposed selection of the candidate block VRF with the miners initially selected based on PoS, theoretically
$$P(forking) \sim 0$$ since only the candidate block is propagated after the winner miner is selected to other nodes or devices.
\end{proof}

\begin{prop}
The VRF PoS consensus is secure against the 51\% attack and the 50\% hash rate.
\end{prop}
\begin{proof}
In the proposed selection of the BFL role, there are two VRFs $distVRF \text{ and } centVRF$ representing the
distributed VRF that each device has and the central VRF that we use only once, respectively.
In traditional PoS, a group that owns 51\% staked cryptocurrency can alter the blockchain in their favor.
In PoW, that is associated with 51\% computational power.
Thus, the probability of attack is $$ P (attack) \propto 51\% \text{ (stake or computational power)}$$
With the proposed approach, 
$$P(attack) \propto \{ P(dVRFOut \sim cVRFOut) \} . (51\% attack)$$ which 
is comparatively lower than without VRF.
\end{proof}

\begin{prop} \label{prop:hybridsolvesmulti}
Hyrbid Signature Scheme resolves the state issue of multilayer MSS.
\end{prop}
In multilayer XMSS trees with $d$ layers, there are $d-1$ multiple signatures \{$\sigma_1,\sigma_2, .. \sigma_{d-1} $\} for each layer.
These signatures are produced when upper-layer trees certify lower-layer
XMSS trees.
For a multilayer MSS with $d$ layers,
the complexity of the multilayer MSS increases as $d$ increases.
However, the proposed signature approach does not require $d$ layers of XMSS for more keys.
Every time a new XMSS tree is created because we run out of keys in the previous XMSS tree.
Therefore, the number of XMSS layers in the new proposed scheme is always equal to $1$, which reduces the complexity of its implementation.

\begin{prop}
The use of XMSS in the proposed hybrid approach is better than XMMS multitree or HSS schemes.
\end{prop}
For a height XMSS scheme $h$, the size of signature $SIG_{size}$ and the performance $P$ can be defined as, 
$$SIG_{size} \propto h$$.
In a hybrid approach, we use the height of the XMSS tree $H_h < h$.

\begin{prop} \label{prop:synchronization}
The proposed Hybrid Signature Approach solves the synchronization issues for XMSS that can occur if XMSS is used alone.
\end{prop}
The proposed hybrid approach uses Dilithium to certify XMSS trees.
Thus, the public key or hashed value of the Dilithium public key $hash(pk_d)$ can be used as the permanent address for the device.
Thus, in this case, many XMSS trees can be generated and certified by Dilithium.
Dilithium keys remain for the entire lifetime of the device.

\begin{prop}[Correctness]
The correctness of the Hybrid Signature Scheme is formulated and proved here.
\end{prop}
A digital signature uses a private key to sign a message which is verified later using a public key and signature.
Both Dilithium and XMSS used in the hybrid scheme are postquantum digital signature algorithms recommended by NIST.
\begin{enumerate}
    \item Keypair: Hybrid SS still uses XMSS predominantly for signing and verifying transactions. 
    For verification, however, the XMSS tree is signed and verified by Dilithium. 
    Thus, 
    $$XMSS(sign/verify(Tx))$$ whereas, 
    $$Dilithium(sign/verify(XMSStree))$$
    \item Hard to break: Both Dilithium and XMSS are post-quantum signature algorithms.
    \item For signing Message, $m\in M$, the XMSS private key $xPk$ is used, whereas for verification, both $dSig_x$ and $xSig$ are verified.
\end{enumerate}
Thus, the hybrid signature scheme is still a signature scheme that can be
used to sign and verify transactions.

\begin{prop} \label{prop:dilithium_enough}
Verification of only the Dilithium Signature of a particular XMSS tree is enough.
\end{prop}
Initially, for a device, we have $k^{th}$ XMSS tree ($X_{tree}^k$), it is public key ($X_{pk}^k$), private key ($X_{sk}^k$) and Dilithium private key ($D_{sk}$).
Then, dilithium is used to sign $X_{pk}^k$ to generate Dilithium Signature as, 
$$D_{sig}^k \leftarrow D_{sign}(D_{sk}, X_{pk}^{k})$$
For a transaction, $tx$ for a device $d_i$ in a communication round $r$, 
it is signed using XMSS tree $X_{tree}^k$ , such that, 
$$ X_{sig}^k \leftarrow X_{sign}(tx, X_{sk}^k)$$
$X_{sig}$ contains the WOTS+ public key, authentication path and signed message.

In addition, the XMSS public key $X_{pk}^k$ is derived from all public keys $WOTS+$.
To verify XMSS, we compute the XMSS root from the authentication path. 
If the initial public key matches the newly computed root value of XMSS, then XMSS is verified i.e.
$$computedXMSS_{root} == actualXMSS_{root}$$

With Hybrid Scheme, unless the Dilithium Signature is compromised, 
verifying the XMSS signature should be equivalent to verifying the signature of Dilithium Signature of that
XMSS tree.
$$Verify(d_{sig}^k) \sim Verify(X_{sig}^k)$$
Thus, verification of only the Dilithium Signature is enough.







\end{document}

%% file: preamble.tex
\usepackage{color}
\definecolor{officegreen}{rgb}{0.0, 0.5, 0.0}

\usepackage{graphicx}
\graphicspath{{./figures/}{./graphics/}{./graphics/logos/}}

\usepackage{graphicx}
\usepackage{algorithm}
\usepackage[noend]{algpseudocode}
\usepackage{amsmath}

\usepackage{breqn}
\usepackage{subcaption}
\usepackage{multirow}
\usepackage{psfrag}
\usepackage{url}
\usepackage{hyperref}
\usepackage{cleveref}
\crefname{assumption}{assumption}{assumptions}
\usepackage{booktabs}
\usepackage{rotating}
\usepackage{colortbl}
\usepackage{paralist}
\usepackage{epstopdf}
\usepackage{nag}
\usepackage{microtype}
\usepackage{siunitx}
\usepackage{nicefrac}
\usepackage{fontawesome}
\usepackage{xcolor}
\usepackage{multicol}
\usepackage{wrapfig}
\usepackage{todonotes}
\usepackage{tablefootnote}
\usepackage{threeparttable}
\usepackage{cite}
\usepackage{lipsum}
\usepackage[english]{babel}
\usepackage[pangram]{blindtext}
\usepackage{tikz}
\usepackage{tabularx}
\usetikzlibrary{fit,positioning,arrows.meta,shapes,arrows}
\tikzset{neuron/.style={circle,thick,fill=black!25,minimum size=17pt,inner sep=0pt},
	input neuron/.style={neuron, draw,thick, fill=gray!30},
	hidden neuron/.style={neuron,fill=white,draw},
	hoz/.style={rotate=-90}}
\usepackage{mathtools}
\usepackage{amsthm}

\newtheorem{thm}{Theorem}[section]

\newtheorem{prop}[thm]{Proposition}
\theoremstyle{definition}
\newtheorem{defn}[thm]{Definition}
\theoremstyle{remark}

\numberwithin{equation}{section}

\usepackage[
    n,
    operators,
    advantage,
    sets,
    adversary,
    landau,
    probability,
    notions,    
    logic,
    ff,
    mm,
    primitives,
    events,
    complexity,
    asymptotics,
    keys]{cryptocode}
    
\usepackage{amsfonts}



%% file: abstract.tex
\begin{abstract}
Post-quantum security is critical 
in the quantum era. 
Quantum computers, along with quantum algorithms, make the standard cryptography based on RSA or ECDSA over FL or Blockchain vulnerable.
The implementation of post-quantum cryptography (PQC) over such systems is poorly understood as PQC is still in its standardization phase.
In this work, we propose a hybrid approach to employ PQC over blockchain-based FL (BFL), where we combine a stateless signature scheme like Dilithium (or Falcon) with a stateful hash-based signature scheme like the extended Merkle Signature Scheme (XMSS).
We propose a linear-based formulaic approach to device role selection mechanisms based on multiple factors to address the performance aspect.
Our holistic approach of utilizing a verifiable random function (VRF) to assist in the blockchain consensus mechanism shows the practicality of the proposed approaches.
The proposed method and extensive experimental results contribute to enhancing the security and performance aspects of BFL systems.

\end{abstract}

%% file: body.tex
\section{Introduction}
Existing blockchain and federated learning (FL) systems currently rely on classical cryptographic methods, such as RSA or ECDSA, to ensure system security. However, the emergence of quantum algorithms such as Shor's algorithm poses a significant threat to the safety of these systems as large-scale quantum computers become more accessible
\cite{dichianoSurveyNISTPQ2021}. 
This necessitates the adoption of post-quantum security measures to mitigate the risks.
Mosca's Theorem provides the urgency and time frame for the preparation for quantum security. 
It states that adopting post-quantum security is mandatory, 
when $$\textit{Security shelf life} + \textit{Migration time} \geq \textit{Collapsetime}$$
where, 
\textit{Security 
shelf life} refers to the duration of time for which the data should be kept secure for;
\textit{Migration time} is the time required to implement quantum security; and
\textit{ collapse time} denotes the time it will take for quantum computing to be available.
\begin{figure}[!h]
    \centering
     \includegraphics[width=0.7\columnwidth]{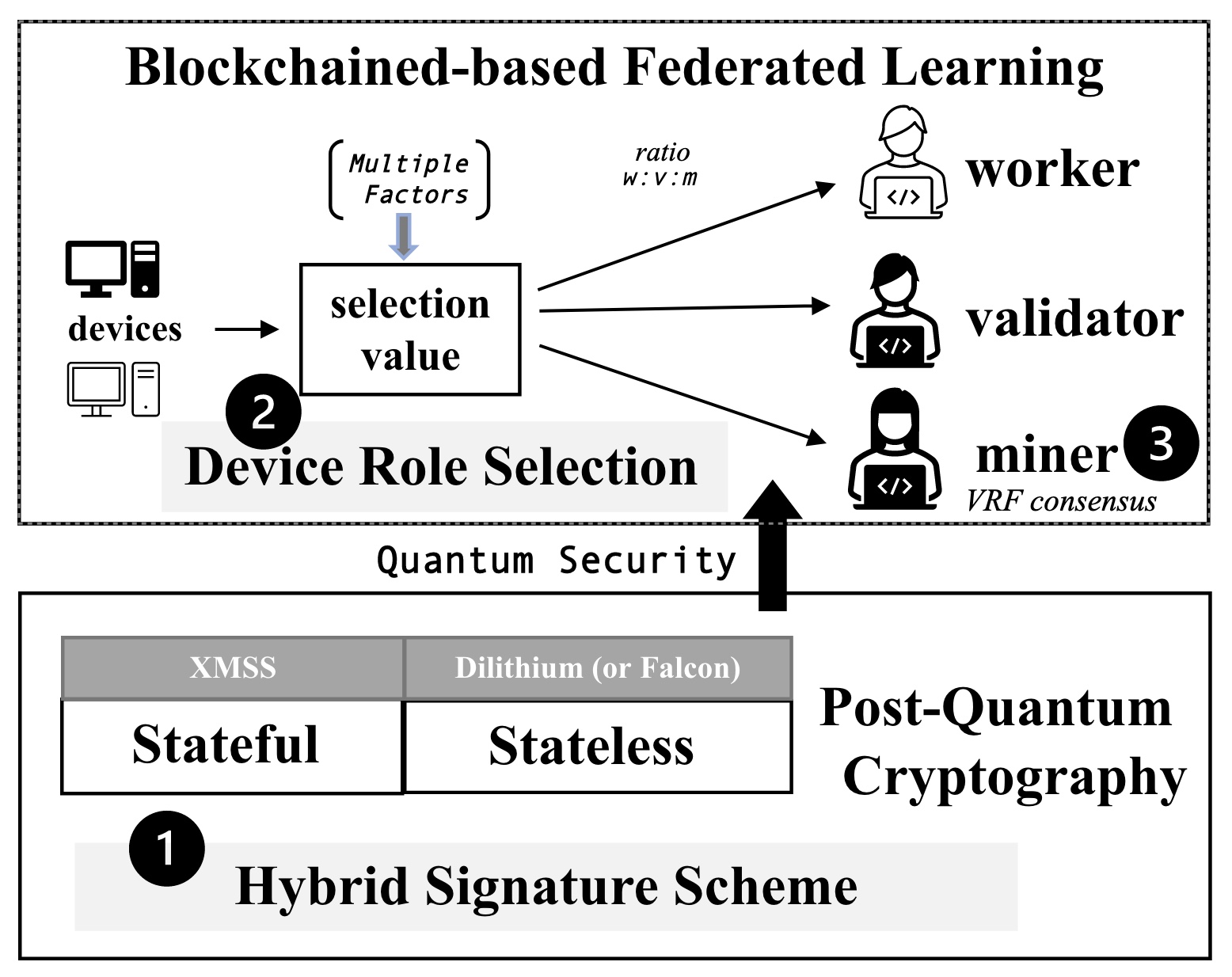}
    \caption{\textcircled{1} A proposed hybrid approach to postquantum secure signature schemes, \textcircled{2} device role selection mechanism, and \textcircled{3} use of VRF for consensus mechanism.}
    \label{fig:intro_image}
\end{figure}
To address the security threats posed by quantum computing, the National Institute of Standards and Technology (NIST) has been leading the standardization process for postquantum cryptography (PQC) \footnote{https://csrc.nist.gov}. 
However, many PQC signature schemes proposed by NIST have a larger signature and public key sizes than classical RSA. 
Moreover, their key generation, signing, and verification performance are slower than that of classical cryptography. 
These PQC schemes are still not extensively studied and explored in the context of their implementation in blockchain and federated learning (FL) systems, making it challenging to understand their crypto-agility.
On the other hand, hash-based signature schemes, particularly stateful ones, offer quantum-proof security and have been well studied \cite{butinHashBasedSignaturesState2017}. They also demonstrate better performance compared to other PQC schemes. Stateful hash-based signature schemes (HBS) rely on the minimal security assumption of hash-based functions, a crucial aspect of their design \cite{hulsingHashbasedSignaturesOutlinea}. However, implementing these schemes can be challenging due to their stateful nature, as the generated keys have limited usage \cite{cooperRecommendationStatefulHashBased2020a}.
Performance is a crucial consideration in addition to security regarding blockchain-based federated learning (BFL) systems \cite{pokhrelFederatedLearningBlockchain2020a}. Several factors can influence the system's performance, including the number of participating devices, the signature schemes deployed, the ratio of the number of roles, the consensus mechanism, the rewarding mechanism, and the computing resources of the devices. As research on Verifiable Random Function (VRF) in blockchain and postquantum VRF continues to evolve \cite{pqvrf}, it is essential to explore how these mechanisms can be integrated into BFL systems to enhance their performance. Furthermore, optimal client selection is an emerging area of interest in federated learning; however, the criteria for selecting clients must be reassessed to realize their potential fully.
This paper focuses on addressing two research challenges:
\begin{enumerate}
\item [${R_1}$] (Security Challenge) 
The first challenge we address is designing an efficient post-quantum cryptography (PQC) framework for blockchain-based federated learning (BFL) systems and ensuring their security against quantum attacks.

\end{enumerate}
Furthermore, as the number of devices, data, and network size grows, more research is needed to improve the performance of federated learning (FL) and BFL systems.

\begin{enumerate}
\item [${R_2}$] (\textit{Performance Bottleneck}) 
The second challenge we tackle relates to the client selection mechanism. We aim to determine the most efficient tool for selecting devices as workers, validators, and miners in a blockchain-based FL environment where these roles exist concurrently. 
Furthermore, we explore how verifiable random function (VRF) can be crucial in improving consensus mechanisms for blockchain-based distributed systems.
\end{enumerate} 

Following the research challenges mentioned above, we make the following key contributions.

\begin{enumerate}
\item We propose a hybrid approach that combines stateless post-quantum secure signature schemes, such as Dilithium or Falcon, with stateful hash-based signature schemes like XMSS. In this approach, XMSS is employed to sign actual transactions, while Dilithium is used to certify the XMSS signature scheme.

\item We introduce a linear model for device role selection in blockchain-based federated learning (FL) systems and device selection in standalone FL systems. Our model utilizes a fuzzy logic formulaic approach, calculating a "selection value" based on multiple factors discussed in detail in our paper.

\item We present a comprehensive approach to using VRF to improve consensus mechanisms in blockchain-based FL systems.
\end{enumerate}

\section{Problem Definition}
In this section, we highlight the potential vulnerabilities of
blockchain and federated learning (FL) systems to adversarial attacks, particularly in the quantum era. 
It emphasizes the necessity of adopting post-quantum security measures for blockchain-based FL systems. 
Additionally, it underscores the importance of developing new role selection mechanisms and enhancing consensus mechanisms.

In the quantum era, adversarial attacks can be launched by adversaries equipped with quantum computers capable of running algorithms like SHOR's algorithm. 
Such attacks can have detrimental consequences, including data privacy breaches and model security. 
Implementing post-quantum security measures can also introduce performance issues that manifest themselves in key generation, signing, and verification processes. 
Furthermore, there is a constant need for performance optimization in any system over time. 
Effectively addressing the trade-off between performance and security is crucial. 
While post-quantum cryptography (PQC) provides quantum security, 
it may also introduce performance bottlenecks due to the characteristics of the signature scheme employed.

\section{Background}
In recent years, NIST has been working towards the standardization
of several PQC schemes to prepare for the quantum era.
Among all candidate schemes, hash-based signature schemes, such as XMSS and LMS, have already been recommended by NIST to be used for early implementations.
Just recently, last year, in 2022, NIST also selected three signature schemes: CRYSTALS-dilithium \footnote{https://pq-crystals.org/dilithium/}, FALCON \footnote{https://falcon-sign.info/}, and
SPHINCS+ \footnote{https://sphincs.org/}.
As integer factorization and discrete logarithm-based cryptography such as RSA and ECDSA are weak against quantum algorithms, 
extensive research work has been carried out regarding the study
and benchmarking of PQC schemes.
But implementation of PQC in
blockchain and FL networks, which is still a much-uncovered area, is crucial for postquantum security.
For a smooth transition from classical security to postquantum security, a better understanding of the practicality and suitability of PQC
is required.
That leads to an important factor that needs to be addressed, termed as 
"\textit{crypto-agility}", 
which refers to the ability of a system to
adapt to new cryptographic schemes quickly
without the whole system being affected \cite{wiesmaierPQCMigrationCryptoAgility2021}.
Current cryptographic schemes used in blockchain and FL systems like RSA and ECDSA for signature are not
quantum-safe, meaning adversaries can easily break them with access to quantum computers.
Shor's algorithm with a quantum computer can provide exponential solutions for integer factorization and discrete logarithm problems.
Thus, public key cryptography signature schemes like the Elliptic Curve Algorithm (secp256kl) used by Bitcoin are
weak to Shor's algorithm
\cite{alkeilanialkadriDeterministicWalletsQuantum2020,sharmaAnalysisCrystalsDilithiumBlockChain2021}. 
Gorver's algorithm can compute new hashes so fast that an entirely new blockchain can be created.
It also impacts symmetric algorithms and hash functions at their current bit-security sizes. 
Blockchain platforms such as Bitcoin and Ethereum use cryptography to generate addresses, for consensus mechanisms, and to protect transactions.
Table \ref{tab:blockchain_networks} shows the current state of the art in cryptography used by most famous blockchain networks.
\begin{table}[!h]
    \centering
    \resizebox{\columnwidth}{!}{%
    \begin{tabular}{|l|c|l|}
    \hline
       Blockchain  &  Cryptography & Consensus\\
       \hline  \hline
         Bitcoin & Secp256kl  & Proof of Work \\
         Ethereum & Secp256kl & Proof of Work $->$ Proof of Stake \\
         Hyperledger Fabric & ECDSA & Consortium Consensus \\
         Algorand & Ed25519 & Pure PoS \\
         R3 Corda & Secp256r1 & Validity \& Uniqueness\\
     \hline
    \end{tabular}}
    \caption{Blockchain Networks}
    \label{tab:blockchain_networks}
\end{table}

\subsection{Extended Merkle Signature Scheme (XMSS)}
XMSS is a stateful hash-based signature scheme.
It requires careful state management of the keys.
A single key to XMSS cannot be used multiple times for security reasons. 
Whereas schemes like Dilithium and Falcon are stateless, meaning
they don't need to update their keys (in contrast to XMSS).
The key generation time of XMSS depends on the height of its tree.
The deeper the tree, the longer it takes to generate the keys.
According to NIST guidelines,
to be considered a full-fledged signature scheme, the scheme must produce at least $2^{64}$ signatures \cite{europeanunionagencyforcybersecurity.PostquantumCryptographyCurrent2021}.
To fulfill the NIST requirement, 
the tree height must be $64$, which considerably increases the key generation time
(almost impractical for real-time Blockchain and FL systems, 
as shown in table \ref{tab:xmss_key_generation_time}). 
However,  stateful HBS schemes like XMSS stand out from other PQC signature schemes because they rely on minimal security assumptions \cite{buchmann2011xmss}.
The security of HBS has been tested and well studied over several years, 
which implies that they are more mature enough than the other schemes.
\begin{table}[]
    \centering
    \resizebox{0.6\columnwidth}{!}{
    \begin{tabular}{|c|c|l|}
    \hline
      \textbf{Height}  &  \textbf{No. of WOTS+ Keys} & \textbf{Time (secs)}\\
      \hline
        2&  4 & 0.045\\
        4& 16 & 0.14 \\
        6 & 64 & 0.57 \\
        8 & 256 & 3.01  \\
        10 & 1,024 & 12.09  \\
        12 & 4,096 & 39.28  \\
        14 & 16,384 & 156.27\\
        16 & 65,536 & 631.06\\ 
        18 & 26, 2144 & 2,611.28\\
        20 & 1,048,576 & 9,823.49\\
        \hline
    \end{tabular} }
    \caption{XMSS key generation time and the number of keys available based on different tree heights.}
    \label{tab:xmss_key_generation_time}
\end{table}
\subsection{Dilithium and Falcon}
Both Dilithium and Falcon are post-quantum secure digital signature schemes.
They are stateless and are both lattice-based.
 Dilithium is a digital signature scheme whose design is based on the "Fiat-Shamir with Aborts" approach. 
It is lattice-based cryptography, depending on the hardness of the lattice problems.
It uses the SHAKE hashing algorithm, which stands for Secure Hash Algorithm. KECCAK was used before, but for round 3, it was replaced with AES \cite{raaviPerformanceCharacterizationPostQuantum2021}. 
Different versions of dilithium according to the security level are Dilithium2, Dilithium3, and Dilithium5, which indicate security levels of 2, 3, and 5, respectively.

Falcon, a post-quantum cryptographic signature algorithm, stands for fast Fourier-lattice-based compact signature scheme over an N-th Degree Truncated Polynomial Ring (NTRU).
\cite{fouqueFalconFastFourierLatticebaseda}. 
It is based on the GPV (Gentry, Peikert, and Vaikuntanathan) theoretical framework for lattice-based signature schemes. 
This framework is instantiated over NTRU lattices using 'fast Fourier Sampling,' a trapdoor sampler.
Falcon scheme can be mathematically described as FALCON = GPV Framework + NTRU Lattices +  Fast Fourier Sampling . 
The Falcon scheme has two variations: Falcon512 and Falcon1024, representing NIST security levels 1 and 5, respectively.
Falcon is based on a complicated problem: a short-integer solution problem (SIS) over NTRU lattices.
Even with quantum computers, there is no efficient algorithm to break it.
Some important highlights of Falcon Scheme include:
use of Gaussian Sampler, which guarantees the security of signatures, usage of NTRU lattices leading to shorter signature sizes(compactness), 
fast implementation because of fast Fourier sampling usage which leads to 1000 signatures/sec on local computers and 5x to 10x faster verification. 
Falcon's time complexity is of the order O(n log n) for degree n, and its key generation uses less than 30 KB of RAM.

\section{Related Work} \label{sec:literature_review}
Of particular relevance to this work are research
on FL, blockchain-based FL, and blockchain in terms of security and selection mechanisms.
We have summarized and compared the closest works
in Table \ref{tab:literatur}
in terms of performance and security aspects. 
From a performance point of view, Pokhrel \textit{ et al.}~\cite{pokhrelFederatedLearningBlockchain2020a} proposed blockchain-based FL for autonomous vehicles. 
They developed a new method for improving communication bottlenecks and
computation delay of blockchain-based FL
by modeling block generation rate and forking using
online convex optimization. 
For incentive mechanisms, 
Xu \textit{et al.} \cite{Rongxin} proposed the Fair-blockchain-based FL framework,
which is a contribution-based incentive mechanism to select highly contributing clients.
But both works \cite{pokhrelFederatedLearningBlockchain2020a, Rongxin}
adopted classic RSA and didn't
analyze the post-quantum security aspects.
Chen \textit{et al.} \cite{chenRobustBlockchainedFederated2021} proposed a robust
validation mechanism to identify malicious nodes and improve blockchain-based FL performance.
The submitted validation mechanism performed better than vanilla FL in the presence of malicious nodes.
However, their device selection is random and security is not quantum-proof.

Among different studies related to the selection mechanism,
Deng \textit{et al.} \cite{dengAUCTIONAutomatedQualityAware2022} proposed the AUCTION framework,
which is capable of evaluating the learning capacity of clients and
thus selecting clients on that basis.
Huang \textit{et al.} \cite{huangStochasticClientSelection2022} proposed a stochastic client selection algorithm
under the assumption of volatile context, which affects local training, while Cho \textit{et al.} \cite{powerofchoice} analyzed biased client selection
for sharp error convergence to improve communication efficiency in heterogeneous environments by selecting clients with higher local loss. 
Xin \textit{et al.} \cite{xinFederatedLearningClient2022} addressed client selection in the heterogeneity of the system and data.
They minimized the overall training time without affecting accuracy.
Their algorithm
adaptively adjusted the number of selected clients based on their performance.
Batool \textit{et al.} \cite{batoolFLMABClientSelection2022} put forward FL-Multi-Auction using Blockchain (FL-MAB) 
that employs a client selection mechanism based on each client's compute and network resources,
together with the quality of local data.
Lai \textit{et al.} \cite{laiOortEfficientFederated} developed participant selection, 
giving priority to clients with quality data and those that can train quickly. 
All these works \cite{dengAUCTIONAutomatedQualityAware2022, huangStochasticClientSelection2022, powerofchoice, xinFederatedLearningClient2022, batoolFLMABClientSelection2022, laiOortEfficientFederated} are only studied for FL systems and without a post-quantum security aspect.
Thus, our work differs from the literature mentioned above as we investigate performance and quantum security aspects for FL and blockchain-based FL systems.

For the post-quantum security aspect, 
 Lyu \textit{et al.} \cite{fairPrivacyPreservingFL} introduced a deep learning framework that preserves decentralized fairness and
privacy.
Peng \textit{et al.} \cite{LatticeBasedFederatedLearning} proposed
a mechanism to maintain privacy for client gradients against quantum attacks by using the New
Hope lattice-based key exchange scheme.
 Similarly, Zuo \textit{et al.} \cite{PrivacyPreservingAggregationFLLattic} also proposed a lattice-based postquantum privacy-preserving aggregation protocol.
 Unlike these works \cite{fairPrivacyPreservingFL, LatticeBasedFederatedLearning, PrivacyPreservingAggregationFLLattic}, our work focuses on the security aspect of blockchain-based FL so that communication between devices cannot be affected by quantum threats.
We use PQC schemes such as Dilithium, Falcon, and XMSS for that.
While most of the above literature focuses on client selection in FL, our work focuses on more
than just client selection, as blockchain-based FL involves more than a single type of client, i.e., multiple roles.
Thus, our work focuses on a selection of not just devices, but instead roles for devices in each communication round for blockchain-based FL systems. 
Also, we consider more than one dependent factor in our device role-selection approach.
\begin{table}[!h]
    \centering
    \resizebox{\columnwidth}{!}{%
    \begin{tabular}{|c|c|c|c|l|}
    \hline
      References  & FL & blockchain-based FL &  Security  & Performance  \\
     \hline
      \cite{pokhrelFederatedLearningBlockchain2020a} & \checkmark & \checkmark & RSA & Efficient Communication \\
      \cite{powerofchoice} & $\checkmark$ &- & - & Local Loss  \\
      \cite{huang2021shapley, laiOortEfficientFederated} &$\checkmark$ & - & - & Data Distribution  \\
      \cite{chenRobustBlockchainedFederated2021} & $\checkmark$ & - & RSA & Validation Mechanism\\
      \cite{Rongxin} & $\checkmark$ & $\checkmark$ & RSA & Incentive Mechanism \\
     \cite{dengAUCTIONAutomatedQualityAware2022} & $\checkmark$ & -  & - & Data Size, Quality, Learning Budget \\
    \cite{xinFederatedLearningClient2022} & $\checkmark$ & - & - & Latency/Cluster Model\\ 
    \cite{batoolFLMABClientSelection2022} &$\checkmark$ & $\checkmark$ & -  & Computer/Network Resource, Data Quality\\
    \cite{LatticeBasedFederatedLearning,PrivacyPreservingAggregationFLLattic} & $\checkmark$ & - & Lattice-Based Encryption & Privacy Preserving \\
      \textbf{This work} & $\checkmark$ & $\checkmark$ & XMSS/Dilithium & Role Selection on Multiple Factors / VRF implementation \\
      \hline
    \end{tabular}}
    \caption{Literature on blockchain-based FL, FL and Client Selection related Security and Performance.}
    \label{tab:literatur}
\end{table}

\section{Relevant Ideas and Theories}
This section includes a description of previous literature that inspired our work.
Our proposed hybrid approach to signature schemes is built on top of concepts used in the Merkle
signature scheme, hypertrees, Goldreich Approach, and SPHINCS schemes planned here.

\subsection{Acronyms}
\begin{table}[h]
    \centering
    \resizebox{0.8\columnwidth}{!}{%
    \begin{tabular}{|l|l|}
     \hline
      Acronyms   &  Meaning \\
     \hline \hline
      PQC & Post Quantum Cryptography \\
      NIST & National Institute of Standards and Technology\\
      OTS & One Time Signature\\
      WOTS & Winternitz One Time Signature\\
      XMSS & eXtended Merkle Signature Scheme\\
      HBS & Hash-Based Signature \\
      MTS & Merkle Tree Signature \\
      MSS & Merkle Signature Scheme \\
      HSS & Hierarchical Signature Scheme \\
      LD-OTS & Lamport-Diffie OTS\\
      $XMSS^{MT}$ & Multi-Tree XMSS\\
      ECDSA & Elliptic Curve Digital Signature Algorithm \\
      RSA & Rivest–Shamir–Adleman\\
      DSA & Digital Signature Algorithm\\      
      \hline
    \end{tabular}}
    \caption{Acronyms and their Meaning}
    \label{tab:acronyms}
\end{table}

\subsection{WOTS+}
W-OTS+, proposed by \cite{hulsingWOTSShorterSignatures2013}, is a modification 
 of WOTS which is used by the XMSS signature scheme.
The main contribution is the introduction of bit-mask XOR in the chaining function.
Also, in contrast to WOTS, the message is divided into $log_2(w)$ bits, not in $w$ parts.
This introduced a trade-off of a better computational cost for larger signatures and key sizes.
In the W-OTS+ scheme, we have the secret key, a certain number of function chains, and the public key, which consists of the final output of the function chain.
W-OTS+ is different from other variants of W-OTS in terms of the mode of iteration used in the construction of function chains. 
It is also one of the most popular OTS schemes.
For verification purposes, a public key can be generated from the signature itself.
Thus, they can fit into signature schemes like XMSS.
With WOTS+, there is no need for a collision-resistant hash function
.
One disadvantage of WOTS+ is that the key generation and signing time are related to each other.
It means there is a trade-off between those two.
Normally, verification might have to do several times, thus in such situations, 
there could be a problem with a slower verification time.

\subsection{Merkle Signature Scheme }
\begin{figure}[!htb]
    \centering
    \includegraphics[width=0.8\columnwidth]{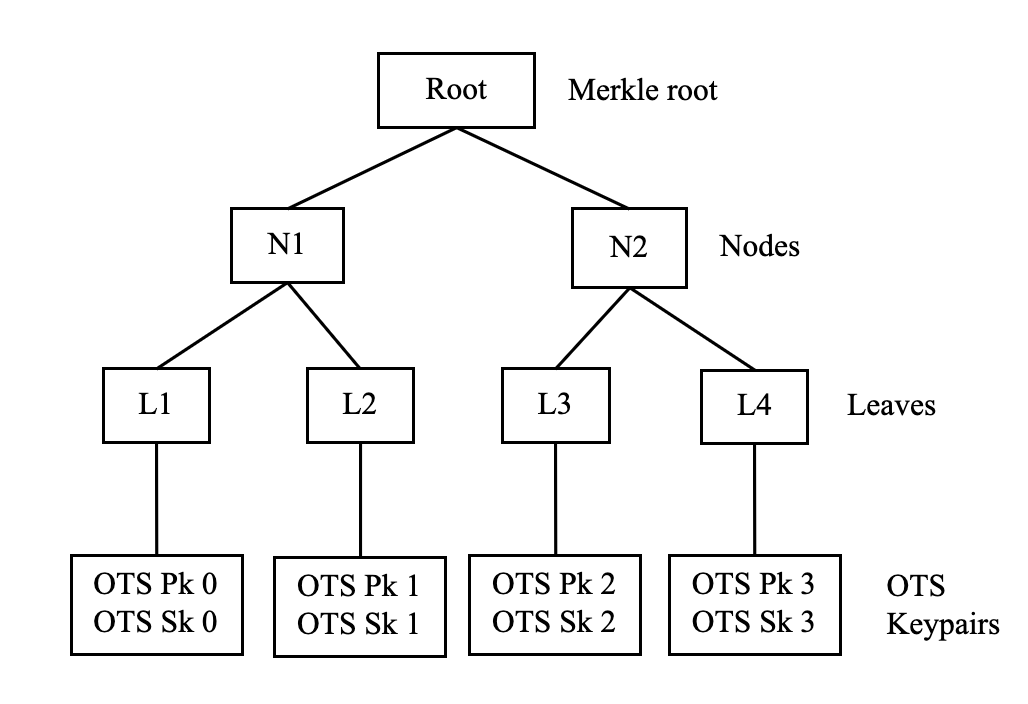}
    \caption{A Merkle Signature Scheme}
    \label{fig:merkle_signature}
\end{figure}

Merkle Signature Scheme (MSS)  is based on merkle trees.
Merkle trees are used to convert, otherwise, one-time-only impractical signature schemes into many-time 
signature schemes with validation required against only one single public key.
An MSS consists of
OTS key pairs at the leaf node of the Merkle tree.
The value at the leaf node is the hash output of the OTS public key.
Each node present in the tree has the hash value of its children's hash values.
This is the same as with all the leaf nodes and intermediate nodes, all the way up to the top,
to obtain the root value.
This root value is the final hash value, which is the public key of the MSS.
Thus, MSS becomes a many times signature scheme using OTS keys.
We still use a particular OTS key to sign a message only once.
But the difference with MSS is that now every message signed by any of the OTS key pairs can be verified against only one public key, which is the root of the Merkle tree.
In MSS, an important concept is the authentication path; that is,
while signing any message, the signature will also have an authentication path.
The authentication path is used to calculate the root value of the Merkle tree.
For example, in figure \ref{fig:merkle_signature}, if we use the $L1$ leaf OTS key to
sign a message, then we have to keep a record of how we can compute the root value of the Merkle tree
for the purpose of verification later.
Therefore, it means that for leaf $L1$, the authentication path will be ${L2, N2}$.
To break it down, from $L1$, to calculate $N1$, we need $L2$. 
After getting the value $N1$, to calculate the root value, we need $N2$.

\subsection{Goldreich (Huge Trees) \& Hyper Trees}
\begin{figure}[!h]
    \centering
    \begin{subfigure}[b]{0.68\columnwidth}
    \centering
    \includegraphics[width=\columnwidth]{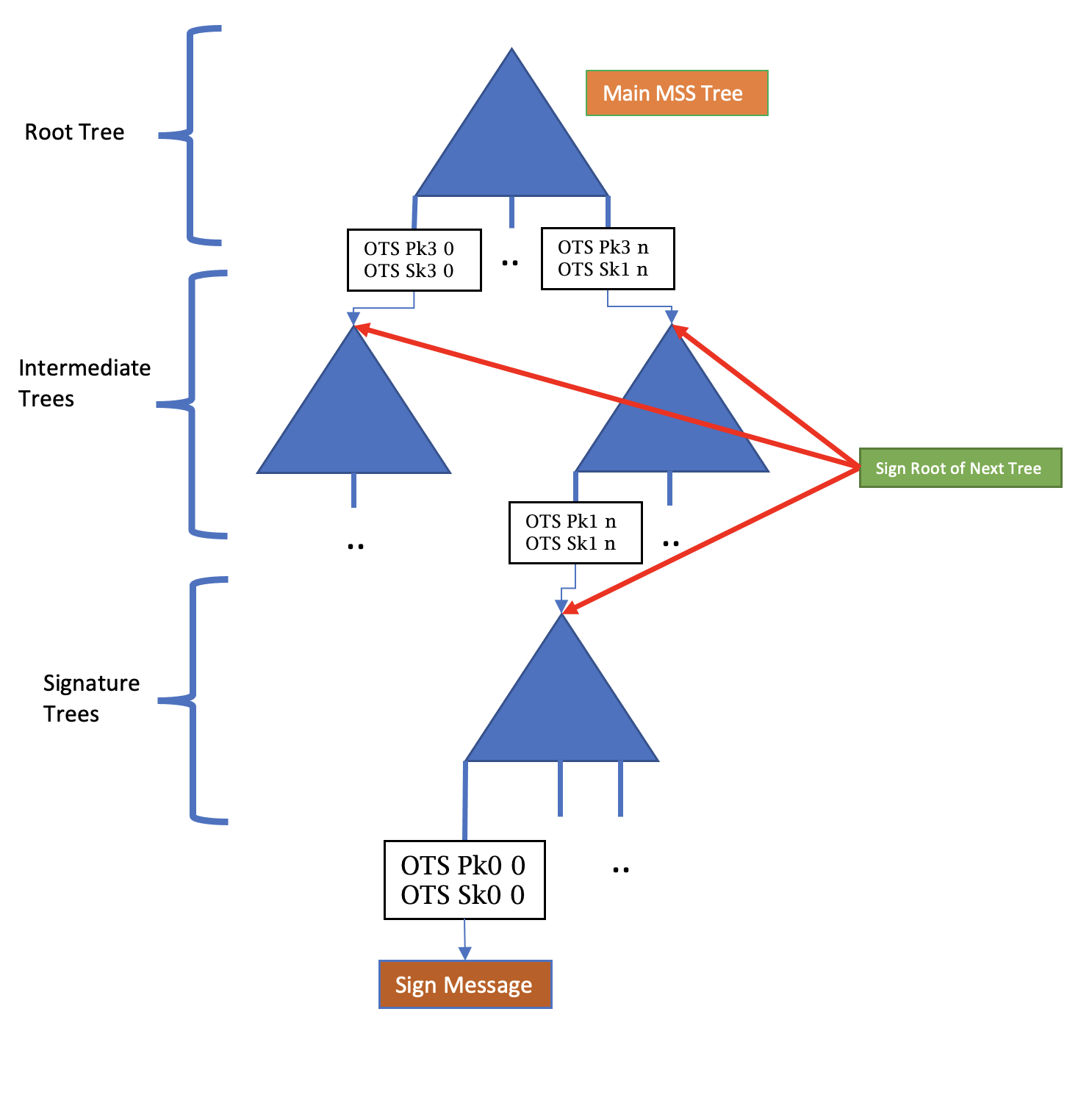}
    \caption{Hyper Tree}
    \label{fig:hypertree}
    \end{subfigure}
    \begin{subfigure}[b]{0.68\columnwidth}
    \centering
    \includegraphics[width=\columnwidth]{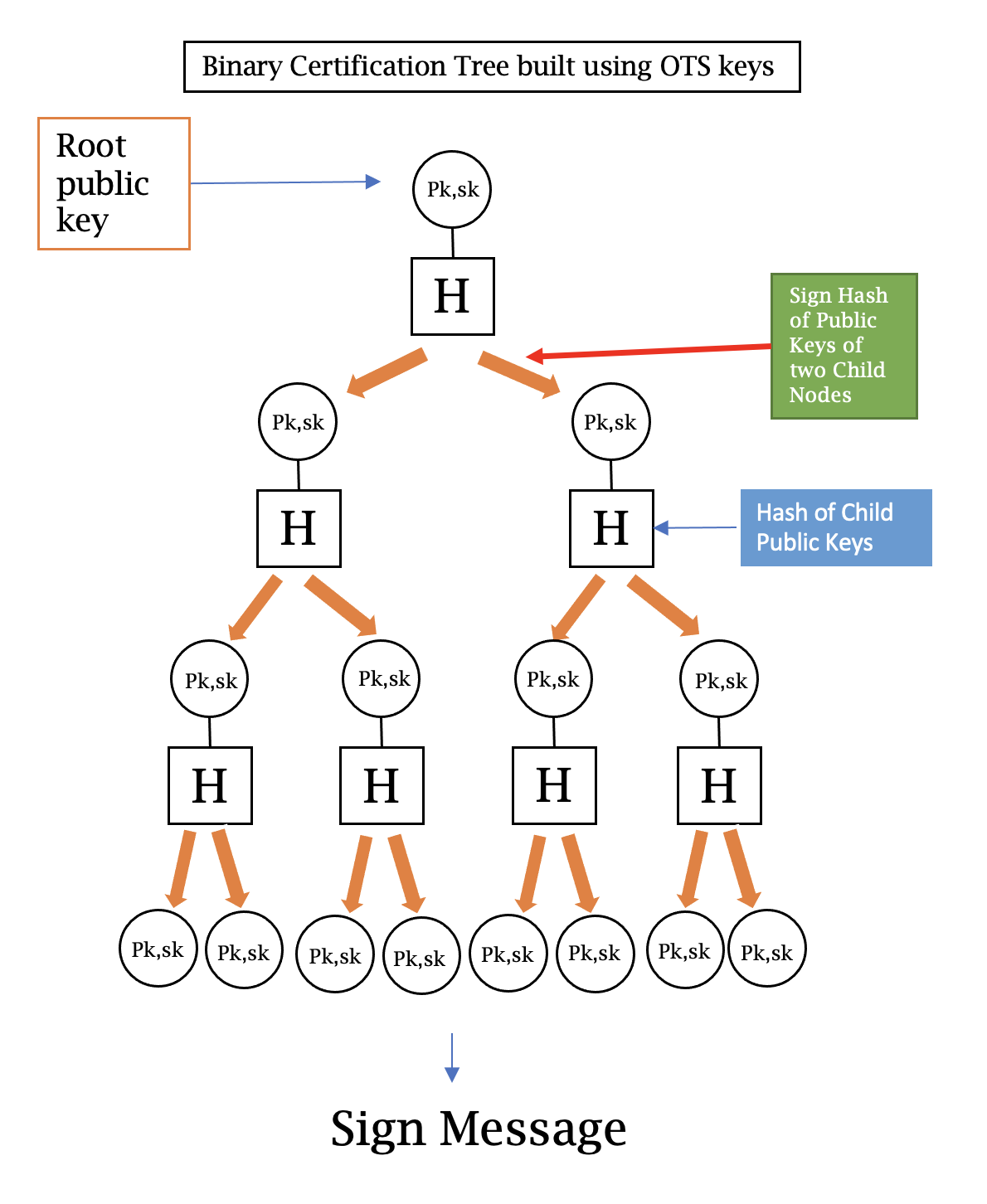}
    \caption{Goldreich Huge Tree}
    \label{fig:goldreich}
    \end{subfigure}
    \caption{Goldreich Huge Tree and Hyper Tree}
    \label{fig:goldreich_tree}
\end{figure}

The two main issues with MSS are memory management and the need to keep records of the keys used, as they cannot be used more than once.
That is why they are referred to as stateful schemes.
To make them stateless, Goldreich defined a typical construction of a huge tree (shown in Figure \ref{fig:goldreich} consisting of
many one-time signature (OTS) keys \cite{goldreichFoundationsCryptography2003}.
 His approach was to have internal nodes also with key pairs in the Merkle tree in contrast to MSS, where the nodes are only the hash of its two child nodes.
 Those OTS keys are used to create a binary certification tree \cite{bernsteinSPHINCSPracticalStateless2015}.
 Nonleaf OTS key pairs are used to sign the hash of its child public keys.
 The leaf OTS keys are used to sign the messages.
 The main root key will be the main public key of the signature scheme.
However, the secret key includes a seed that is used to pseudorandomly generate all the OTS keys of the tree.
 With an extremely huge tree, the probability of the same OTS key pairs being randomly chosen becomes negligible, with
no need to keep the state of which key pair was used.
This definitely addresses the stateful nature of hash-based signatures, but with the cost of bad performance.
 Thus, the main issue with the Goldreich approach is its signature size.
 With the OTS parameter w = 16 and the sha256 hash function, 
 the signature size could increase to 1.65 MB, 
 which is definitely not practical \cite{castelnoviGraftingTreesFault2018a}.
The other type of tree construction called "hyper trees" (shown in Figure \ref{fig:hypertree}) includes variants of XMSS called XMSS-MT
which consists of many layers of XMSS trees,
that is, trees of Merkle trees \cite{hulsingHashbasedSignaturesOutlinea}.
The benefits of using hypertrees include improved key pair generation time compared to XMSS and
only generation of the main root XMSS tree at the beginning.
Other XMSS trees within XMSS-MT are generated as required.
In XMSS-MT, intermediate- and upper-layer XMSS trees are called 'certification trees' which are used to sign lower-layer XMSS trees.
The bottom layer, called 'signature trees', is used to sign the messages.
XMSS-MT definitely solves some problems associated with XMSS, but the problem is with Merkle tree traversal to root calculation and authentication path computation.
Also, with too many XMSS trees, the complexity of node state management becomes too hard,
which eventually leads to an increase in signature sizes as each XMSS certification tree will have produced a signature
while signing the XMSS tree below.
All of these signatures are included in the main signature, which is needed during the verification process.

\subsection{SPHINCS/SPHINCS+}
SPHINCS and SPHINCS+ are other hash-based signatures that are stateless \cite{bernsteinSPHINCSPracticalStateless2015}.
Their construction is also based on the use of hypertrees.
For a total height of $h \in N$ such that $h$ is divisible by $d$, the hyper tree of SPHINCS
consists of $d$ layers of trees, with each tree having a height of $h/d$.
However, leaf nodes that have key pairs to sign messages are few time signatures (FTS) schemes.
Thus, in SPHINCS or SPHINCS+, the main idea is to use a hypertree consisting of a Merkle tree signature to authenticate FTS key pairs \cite{bernstein2019sphincs+}.
FTS schemes can be used to sign multiple messages a limited number of times, as FTS key pairs can produce a small number of signatures.
This is how it is a stateless hash-based signature scheme in SPHINCS.
With SPHINCS/SPHINCS+ we can use them in a stateless manner.
The trade-off is in terms of their key sizes, signature sizes, etc.
Therefore, we still need better solutions to the problem.

\subsection{Stateful Management Approaches}
 McGrew \textit{et. al} \cite{mcgrewStateManagementHashBased2016} provided analysis of state management in N-time hash-based signature schemes.
They proposed the idea of using volatile and non-volatile storage for hierarchical signature schemes.
The idea is to use nonvolatile memory for upper layer trees, whereas
volatile memory is used for the bottom layer, which is used to sign messages.
Also, to protect unintentional copies of the state of private keys, 
they considered a hybrid stateless/stateful scheme.
Their approach solves stateful management to some extent.
But the problem with the approach is increased signature size, and, also,
their recommendations to use SPHINCS with XMSS/LMS are all still few-time hash-based signature schemes.
The use of volatile and nonvolatile memory can still be useful depending on the context and scenarios.
But using SPHINCS along with XMSS will make the size of Merkle trees being used extremely huge,
beyond the scope of practicality.
And SPHINCS is already slow and has bigger key sizes.
Thus, we need better implementation along with a method that can jointly address the issues of state management and computational overhead.

\subsection{Foundations for Device Role Selection}
\subsubsection{Device Roles in Blockchain and FL systems}
In any FL environment, different devices or clients participating play different roles, such as training, validating, aggregating, etc.
Due to many factors like access to computational power, type of data available to the device,
trustworthiness, etc., not all devices can equally perform in the same ways.
Thus, in an ideal situation, we assume, there is no latency, no connectivity issues, etc.
However, this is far from being accurate and therefore needs a better approach to how
the proper role can be assigned to the right device or client in the network depending
upon the circumstances and the capability and trustworthiness of the device.
In a blockchain network setting, especially in Bitcoin, not everyone can be a miner.
Similarly, in Ethereum, only users with the highest stakes can vote for the final block
to be appended.
Also, just randomly selecting any device for a particular role might not be efficient
or practical in the long run.
Thus, client selection, as a first step itself, is really crucial in determining
the overall success of Fl.
Some related works on the selection mechanism have been mentioned in Section \ref{sec:literature_review}.
All studies focus on a selection of clients.
However, a client can perform a different role in any particular communication round.
Thus, an approach needs to address role selection as well, depending on
different reliable factors that directly impact client performance.

\subsubsection{Verifiable Random Function} \label{vrf}
A verifiable random function (VRF) is a function that produces output
and proof after taking input with its secret key.
Anyone can verify that the output was actually pseudo-randomly generated.
The difference from the normal random function with VRF is that
it can be verified that the output was produced correctly.
The output of VRF has properties of verifiability, true uniqueness, pseudorandomness, and collision resistance
\cite{micali1999verifiable}.
VRF is used in a random selection of optimized nodes with varying times for model exchange.
With the help of blockchain and VRF, distributed averaging of
local model updates can be done to achieve a truly
decentralized FL.
Thus, VRFs can be used in decentralized systems to generate a truly random output in response to any query.
One of the crucial aspects of this function is that it can provide trust among participants with proof
that the function is not manipulated.

\noindent \textbf{VRF:} 
VRF includes functions such as key generation, a hashing function, a proof function, and verification.
\begin{itemize}
\item \textit{Key Generation}: Generate the public key of VRF ($vrf_{pk}$) and the private key ($vrf_{sk}$).
    \item \textit{VRF Hash Function}: This function generates the output beta for a given secret key and alpha string ($alpha\_string$).
     $$beta\_string = VRF_{hash}(vrf_{sk}, alpha\_string)$$
    \item \textit{VRF Prove Function}: This function generates the proof $proof$. 
    This proof is used to verify that the output was generated correctly.
    $$proof = VRF_{prove} (vrf_{sk}, alpha\_string)  $$
    \item \textit{VRF Verification Function}: Any verifier with a public key can verify that $beta\_string$ is the correct output from the corresponding secret key $vrf_{sk}$ and the input $alpha\_string$.
    $$result = VRF_{verify}(vrf_{pk}, alpha\_string, proof)$$
\end{itemize}

In contrast to VRF, normal random functions are primarily for simulation and modeling purposes and are not secure enough.
Such functions in Python include random choice, etc. 
Mostly in FL, 
workers are selected in different rounds to do the training locally.
Whereas, a central server will decide or calculate an average of the trained parameters.
The selection process for different roles in particular communication rounds is really crucial in overall FL.
Selection with a normal random function can be manipulated and does not provide information about
how randomness is calculated.
The random function can be manipulated, and thus the node that is going to be selected for
a particular role can be attacked.
This approach has many flaws in the case of the presence of malicious workers.
In the distributed FL approach proposed in most of the literature, 
a committee is chosen to validate the locally trained models.
Most of the time, this is just a random selection.
However, choosing validators to form a committee might be unfair, or the other excluded participants might disagree with the decision.
Here, VRF can play an important role.
It can be used to choose committee members.
Blockchain like Algorand 
uses VRF for cryptographic sortition.
In addition, another focus of ours in using VRF is on consensus mechanisms to assist or replace consensus mechanisms like proof of work or proof of stake.

\section{Proposed Approach}
\begin{figure*}
    \centering
    \includegraphics[width=0.9\textwidth]{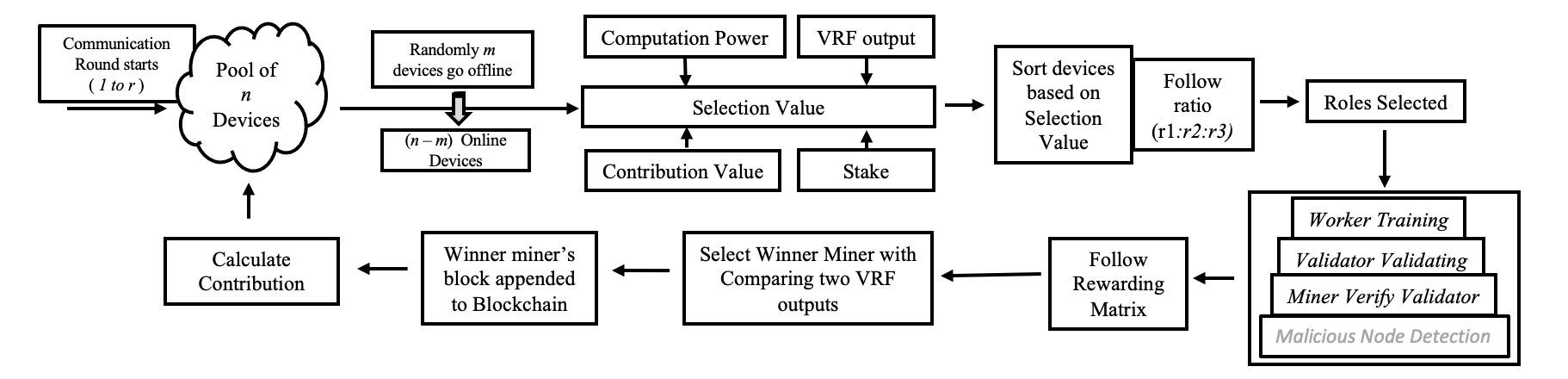}
    \caption{Overall role selection steps.}
    \label{fig:bflsteps}
\end{figure*}
\begin{figure*}
    \centering
    \includegraphics[scale=0.258]{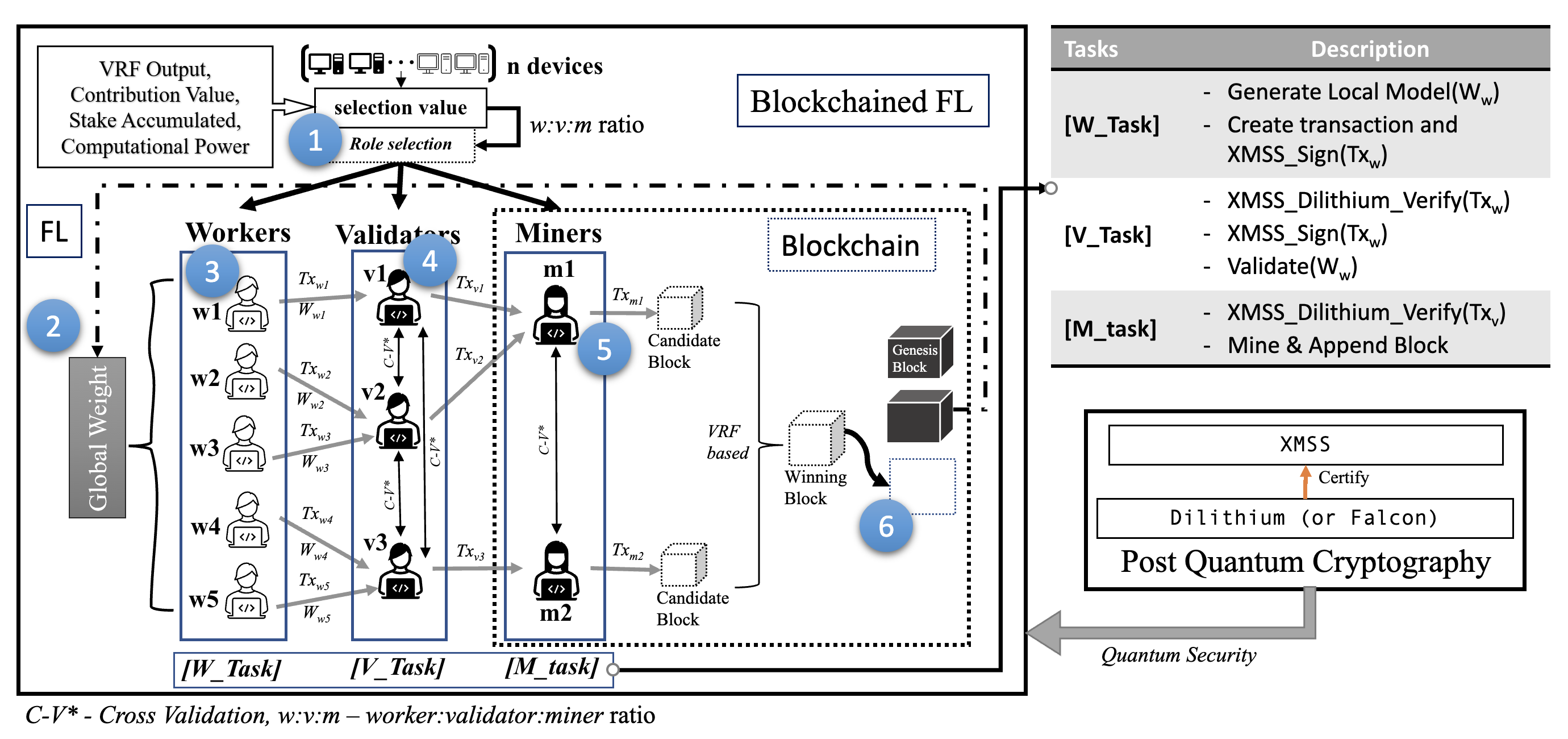}
    \caption{\textbf{Proposed Approaches}: \textcircled{1} Device role selection, \textcircled{2} Global weight used by workers, \textcircled{3} Workers train model, \textcircled{4} Validators validate local models from workers, \textcircled{5} Miners validate transactions from validators and \textcircled{6} newly mined block selected based on VRF is appended to the blockchain. }
    \label{fig:overall}
\end{figure*}

This section presents a detailed overview and description of our proposed approaches.
The clear view, also shown as a representation in figure \ref{fig:intro_image}, is elaborately depicted in figure \ref{fig:overall}.
It shows how we propose to
implement a hybrid approach to two postquantum signature schemes (dilithium and XMSS)
for the system's security and
deploy device role selection based on multiple factors.
The consensus mechanism is assisted by
VRF.
The selection of the winning candidate block is based on VRF comparison. 
As shown in Figure \ref{fig:proposed_hybrid_pqc}, 
first, the devices are assigned a particular role according to their specific characteristics \textcircled{1}.
After retrieving global model weights \textcircled{2}, 
devices with worker role \textcircled{3} train the local model using global model weights.
The workers then send their signed transactions to the
validators to validate the locally
trained model parameters \textcircled{4}. 
After that, the validators sign the transactions and send them to the miner to add the transactions into blocks \textcircled{5}.
Finally, the candidate block is selected from different miners according to the VRF comparison \textcircled{6}.

\subsection{Hybrid Signature Approach} 

\label{subsection:proposed_signature}
To address concerns about the security and practicality of PQC schemes in the quantum era, 
the proposed hybrid signature method
presents the idea of using two PQC signature schemes together. 
The algorithm for the hybrid signature approach is presented in Algorithm \ref{alg:hybridsignature},
while the notation and its meaning are mentioned in Table \ref{tab:notations_used_in_algorithm_hybrid}.
For implementation, in this work, Dilithium (or Falcon) and XMSS are chosen.
In our framework, Dilithium is used to certify the XMSS tree, whereas
XMSS is used to sign transactions.
Along with the signature, a record of the tree number is also kept, 
which is needed in the verification process for the Dilithium of that specific XMSS tree.
To sign the transactions, individual pairs of WOTS + keys are used.
Verification involves verifying the dilithium signature of the XMSS tree as well as the
WOTS+ signature.


\begin{algorithm}
    \caption{Proposed Hybrid Signature Approach}
    \label{alg:hybridsignature}
    \begin{algorithmic}[1]
    \Procedure{\textcolor{blue}{$\kgen$}}{$X, D$}
        \For{each device}
            \State Initialization: $treeNo = 0$, dSignatures = []
            \Procedure{\textcolor{blue}{$\kgen$ D}}{}
                \State $(pk_d, sk_d) \leftarrow D_{keygen}$ 
            \EndProcedure
            \Procedure{\textcolor{blue}{$\kgen$ XMSS}}{$w, h$}
                \For{1 to h}
                {
                 \Procedure{\textcolor{blue}{$\kgen$ WOTS+}}{random.seed}
                \State Generate WOTS+ key pairs.
                \EndProcedure
                }
              \EndFor
            \State Increment $treeNo$ as $treeNo  \mathrel{+}=1$
            \State Create XMSS tree $X_i$ from $2^h$ WOTS+ keys.
            \State Sign XMSS as, $D^i_{s} \Leftarrow D_{sign}(X_i^{pk})$
            \State Append $D^i_s$ to $dSignatures.$
            \EndProcedure
        \EndFor
    \EndProcedure
\Procedure{\textcolor{blue}{signTXs}}{$msg$}
\If{unused $w$ keys available in $X_i$}
    \State $msgDigest \leftarrow hash(msg)$
    \Procedure{\textcolor{blue}{signTXs}}{msgDigest} 
    \State $keyIndex = lastKeyIndex + 1$
    \State Retrieve WOTS keys $(pk_w, sk_w)$ using $keyIndex$
    \State $sig_w \leftarrow WOTS+Sign(msgDigest)$ 
    \State Record $auth$ path nodes.
    \State Return $\{keyIndex, sig_w, pk_w, auth\}$
    \EndProcedure
    \Else 
    \State Create new XMSS tree, $X_{i+1}$
    \State Sign the tree, $D^{i+1}_{s} \Leftarrow D_{sign}(X_{i+1}^{pk})$
    \State Append this new $D^{i+1}_{s}$ to  $dSignatures$.
    \EndIf
    
\EndProcedure
\Procedure{\textcolor{blue}{VERIFY}}{signatures}
    \Procedure{\textcolor{blue}{$X_{verify}$}}{$msg,sig_w,mRoot$} 
    \State $Encode(msg)$
    \State {}keyIndex, sig, otspublic, auth
    \State $cRoot \leftarrow ComputeRoot(auth)$
    \State Verify($mRoot == cRoot$).
    \State Also $wotsVerify(encodedMsg, sig_w)$.
    \EndProcedure
    \Procedure{\textcolor{blue}{$D_{verify}$}}{$msg, deviceId, treeNo$} 
    \State Retrieve $D_s^i$ of $X_i$, where $i = treeNo$
    \State Retrieve public key $pk_d$ using $deviceId$.
    \State Then, $D_{verify}$ $dSignature  dPublickey$.
    \EndProcedure
\EndProcedure
\end{algorithmic}
\end{algorithm}

\begin{table}[!htb]
 \caption{Notations used in Algorithm \ref{alg:hybridsignature}}
    \label{tab:notations_used_in_algorithm_hybrid}
    \centering
     \resizebox{0.6\columnwidth}{!}{%
    \begin{tabular}{|c|l|}
    \hline
     Symbol    &  Meaning \\
     \hline
      $pk_d, sk_d$   & Dilithium Public and Private Key \\
      $pk_w, sk_w$ & WOTS+ Public and Private key \\
      $D_{keygen}$ & Dilithium Key Generation \\
      D & Dilithium Signature Scheme \\
      X & Extended Merkle Signature Scheme \\
      $X_i$ & $i^{th}$ XMSS Tree \\
      $D^i_{s}$ & $i^{th}$ Dilithium Signature \\
      $D_{sign}()$ & Sign using Dilithium \\
      $X_{sign}()$ & Sign using XMSS \\
      $XTree_c$ & Current XMSS Tree \\
      $XTree$ & XMSS Tree \\
      $w$-keys & WOTS+ Keys \\
      $sig_w$ & WOTS+ Signature \\
      $auth$ & Authentication Path \\
      $X_{verify}()$ & XMSS Verification \\
      $D_{verify}()$ & Dilithium Verification\\
      $X_i^{pk}$ & Public Key of $i^{th}$ XMSS Tree \\
       \hline
    \end{tabular}}

\end{table}

\begin{figure}[!ht]
    \centering
    \begin{subfigure}[]{\columnwidth}
    \centering
    \includegraphics[width=0.9\columnwidth]{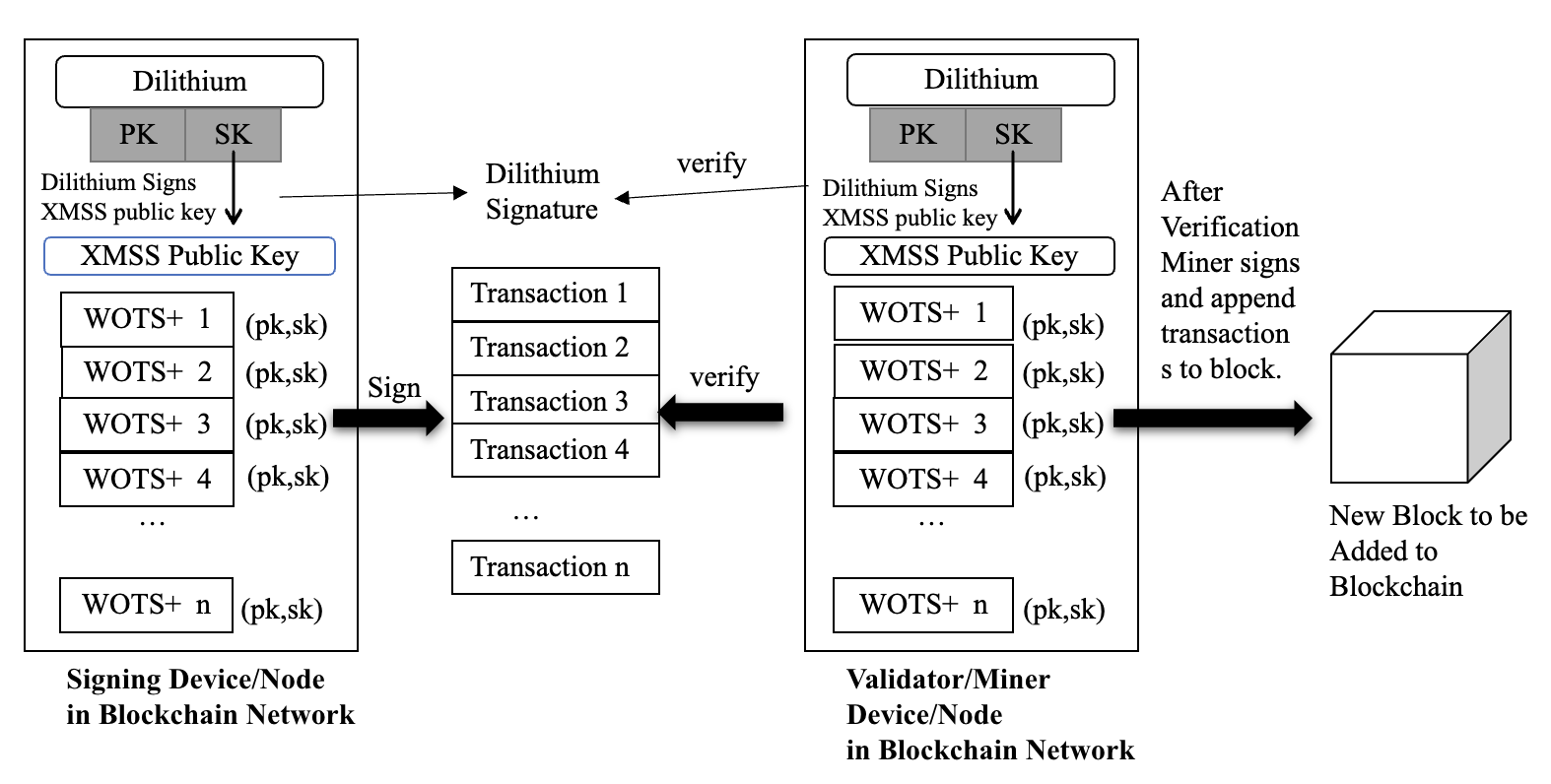}
    \caption{Cryptography view at device level}
    \label{fig:proposed_hybrid}
    \end{subfigure}
    \begin{subfigure}[]{\columnwidth}
    \centering
    \includegraphics[width=0.9\columnwidth]{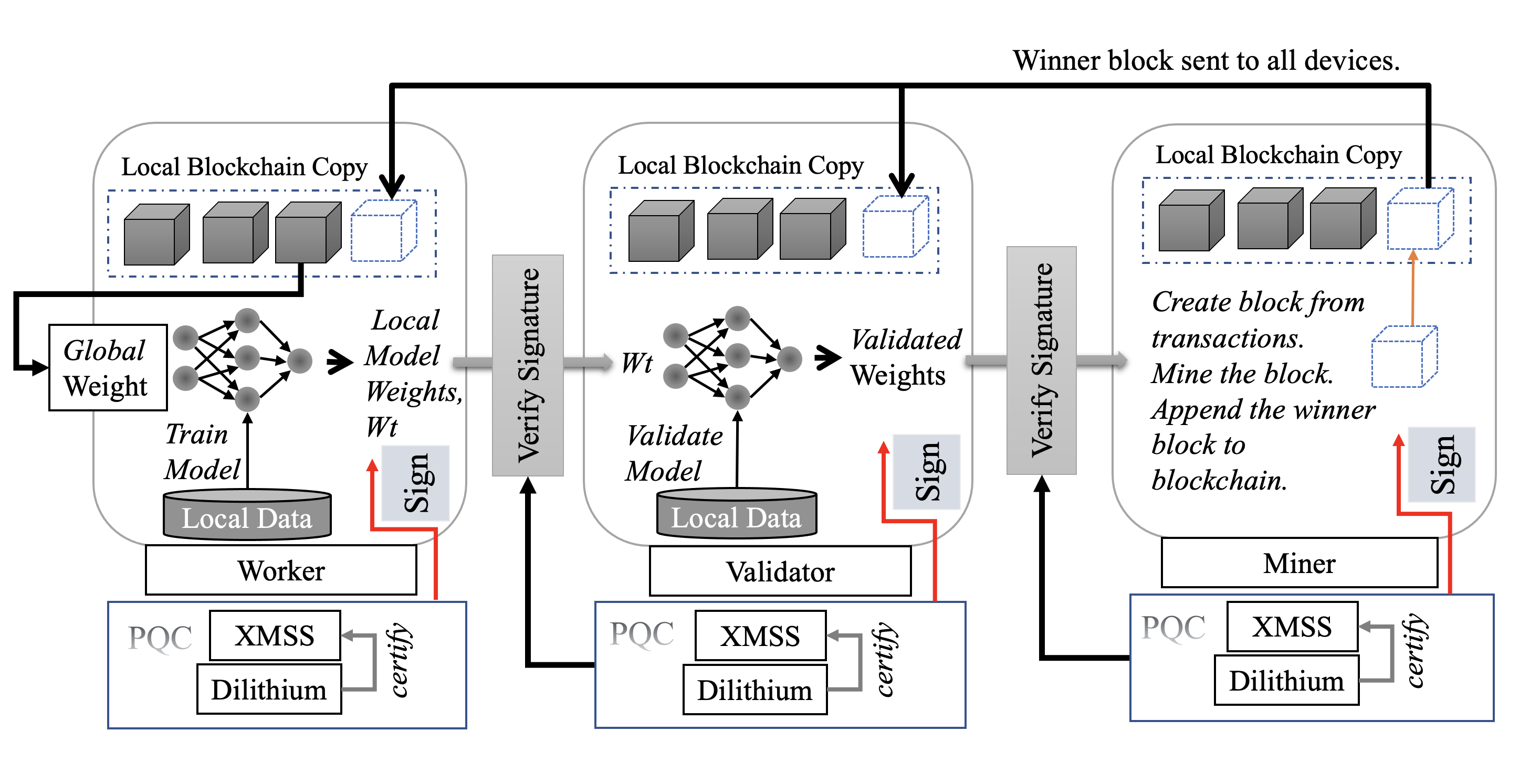}
    \caption{Interaction between Roles}
    \label{fig:interaction_hybrid_pqc}
    \end{subfigure}
     \caption{Proposed Hybrid PQC Signature Approach}
    \label{fig:proposed_hybrid_pqc}
\end{figure}

In Figure \ref{fig:proposed_hybrid}, 
we demonstrate the PQC view of the proposed hybrid signature approach
over blockchain-based FL at the device level. 
Initially, workers sign the transactions with their XMSS keys and
send them for validation to the validators.
After validation, the validators use their own XMSS to sign the validated
transactions and send them to the miners for verification.
After verification, miners add the transactions to their blocks.
Meanwhile, each verification process also involves verifying the dilithium signatures. 
Figure \ref{fig:interaction_hybrid_pqc} demonstrates
the flow of the Hybrid Scheme between the worker, the validater and the miner.
It shows the signing/verification procedure
between either worker and validator or validator and miner.

The proposed approach to the signature scheme is still quantum secure
as both Dilithium and XMSS are quantum secure.

The proposed hybrid scheme has a
$$triple(KGen(D,X), X_{sign}(msg),  (D_{verify}(D_s) $$
and  $$X_{verify}(X_s)))$$
where $KGen(D, X)$ refers to key generation by Dilithium and XMSS, $X_{sign}(msg)$ refers to signing transactions using XMSS, and verification involves verification of XMSS signatures of transactions and
verification of the dilithium signature of the XMSS public key.
$ D_{verify}(D_s)$ refers to the verification of the dilithium signature of the XMSS public key and
$ X_{verify}(X_s)$ refers to the verification of the XMSS signatures.
So, we can write, 
\begin{align*}
& \forall (dKeys, xKeys) \leftarrow \text{KeyGen}(1^{\lambda_x, \lambda_d}) \\
& \forall (msg \in M): [V(D_{verify}(pk_d, pk_x, d_\delta), \\
& X_{verify}(msg, pk_x, x_\delta)) = 1]
\end{align*}
where $\lambda_d, \lambda_x$ are security parameters for Dilithium and XMSS, respectively,
$pk_d$ is Dilithium public key, $pk_x$ is XMSS public key, $dKeys$ and $xKeys$ are Dilithium and XMSS keys respectively. 
This hybrid approach uses Dilithum5 (or Falcon1024), which has a security level of 5 according to NIST's postquantum security categories.
For the security of a digital signature, an attacker should not be able to compute a signature $\sigma_A$ for any message M such that $$ verification(\sigma_A, M) == valid$$ without prior knowledge.
Let us suppose that an attacker with a quantum computer takes
$tD$ time to break the Dilithium signature to retrieve the private dilithium key and $tX$ time to forge XMSS signatures.
Thus, to break a hybrid scheme, an attacker will need at least ($tX$ or $tD$) time, whichever is greater
if the attacker is trying to break the scheme in parallel. 
Also, the attacker will need to break both signature schemes instead of just one.
Thus, the security level $HS_\lambda$ of the hybrid scheme will be $$ HS_\lambda \geq X_\lambda + D_\lambda$$
where $X_\lambda$ and $D_\lambda$ are security levels for XMSS and Dilithium
respectively.
Therefore, the proposed hybrid signature scheme is more secure than just one signature scheme.

\subsection{Device Role Selection Mechanism}
\begin{figure}[H]
    \centering
    \includegraphics[width=0.5\columnwidth]{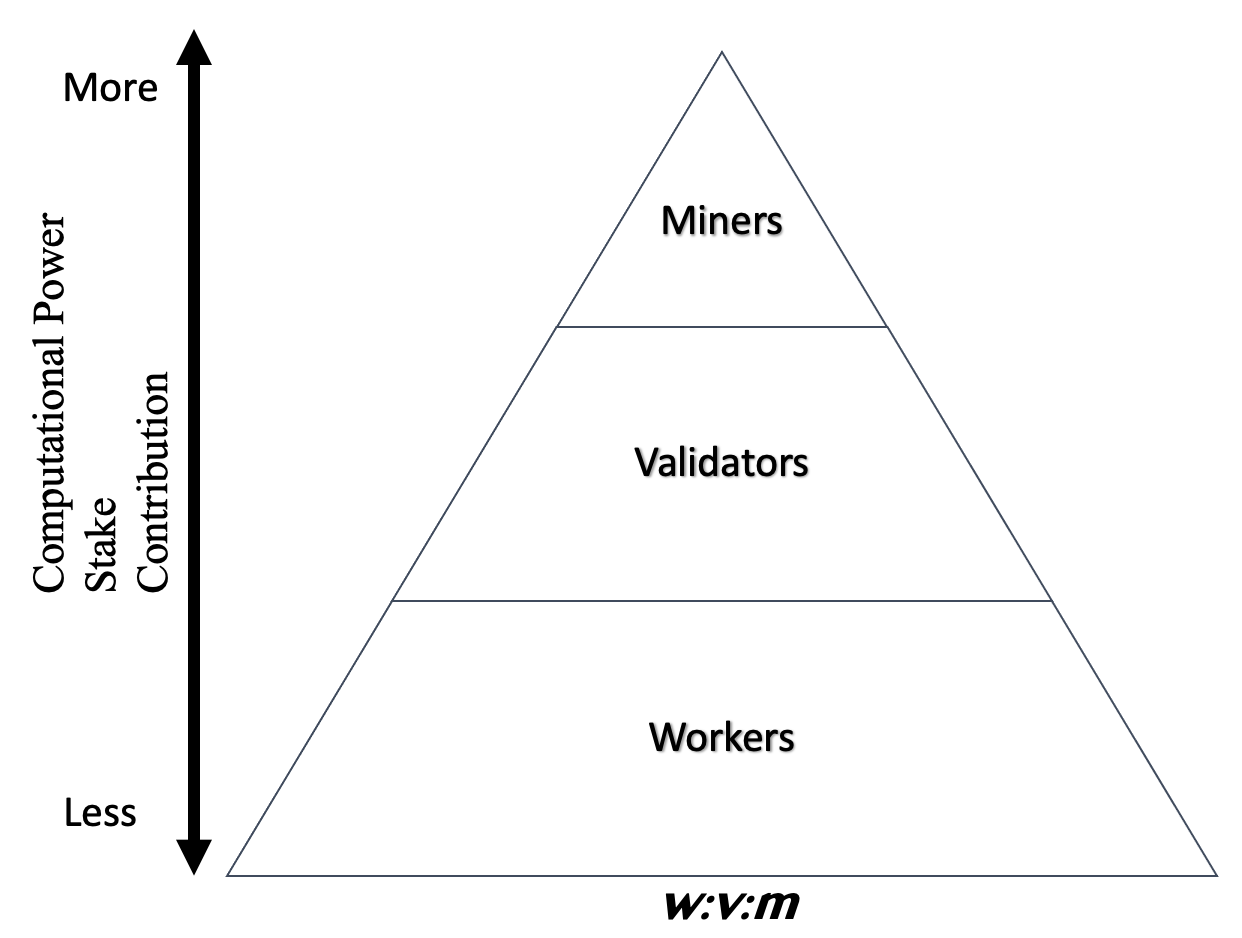}
    \caption{Roles in BFL}
    \label{fig:ecosystem}
\end{figure}

In the real world, 
not every device participating in blockchain-based FL will have similar features
in terms of different aspects such as computational power,
test and training data size,
their contribution, accumulated stake, etc.
In addition to that, 
some devices may be inaccessible due to
network or power failure.
Thus, not every device may be fit to perform the role of a worker, a validator, or a miner under different circumstances at a specific round in a blockchain-based FL scenario.
In terms of blockchain, miners and clients work together
where miners mine the blocks and
clients create transactions.
In the FL scenario, workers are the only clients needed to be selected.
Whatever the scenario, all devices are different.
Studies have been carried out toward client selection but are primarily based only on one or fewer specific
features, for example, gradient-based, local loss-based, etc.
In contrast to that, we emphasize the need for a robust client selection mechanism that can incorporate many other
features of devices.
Thus, we propose a new device selection mechanism to determine
the best role of the suitable device in each communication round autonomously.
Our proposed method calculates a selection value based on different factors, 
which is explained in Algorithm \ref{alg:role_mechanism}.
Basically, the role of a device is selected based on features like computational power,
total stake accumulated throughout previous communication rounds, 
VRF computation, contribution value, learning capability, and difference in the distribution of the data set.
However, these factors and how they are used can be further customized
to make it even more robust and dynamic.\footnote{For the implementation, we have chosen partition ratio as 5:2:1.
We assume that the number of workers will always be greater than at least twice the number of validators and the number of validators around two times at least that of miners.} We have the following assumptions for the mechanism, which is also depicted in Fig. \ref{fig:ecosystem}.

\textbf{Assumption 1:}
For simplicity, we assume that the devices will not go offline after their role is selected. 
Thus, the actively involved devices will work until the end of each round of communication.

\textbf{Assumption 2:}
For the implementation, we have chosen the partition ratio as 5:2:1.
    We require at least 10 devices (there may be more) online to satisfy our minimum partition ratio, which decides the number of workers, validators, and miners in each communication round.
    Thus, we assume that the number of workers will always be greater than at least two times the number of validators and the number of validators around 2 times at least that of miners.

\textbf{Assumption 3:}
Miners are assumed to be the devices that require the highest computational power. 
In the long run, they are also the devices with the most accumulated stakes.

\begin{defn}[Workflow in BFL with Role Selection]
For devices $n$ in $D=\{d_1, d_2, d_3, ....., d_n\}$, initially,  miners = \{\}, workers = \{\}, validators = \{\}, selection values, $sv$ = \{\}.
For each communication round, $r > 0$, the steps involved are:\\
Selection value: The selection value is calculated based on stake $st$,  $vrf$, contribution value $cv$, and computation power $cp$. Thus, $$sv = \{v_1, v_2, ...., v_n\}$$ 
Role selection: The role is selected according to the sorted $sv$ as, in the ratio $$r1:r2:r3$$
\end{defn}

\begin{algorithm}[!h]
  \caption{Proposed Device Role Selection Mechanism} \label{alg:role_mechanism}
  \begin{algorithmic}[1]
  \State INPUT: $n$ devices, device list $D = \{d1, d2, .. , dn\}$
  \State OUTPUT: Role Selected.
  \Procedure{\textcolor{blue}{preparation}}{}
  \For{every device $d$ in D}
      \State Initialize  $sv = 0$.
      \State Compute $vrf$.
      \State Retrieve $s$, $cp$ , $sh$, $cd$, $ls$ and $wd$. 
  \State  Calculate $sv$ using Equation \ref{eqn:role_formula}

  \State Sort $D$ according to $sv$.
  \State Partition sorted $D$ in the ratio of $rw:rv:rm$ by calculating each value $w$, $v$ and $m$  as by using 
  $value = (ratio / sum) * n $
    where, $sum = rw+rv+rm$, $ \text{ratio} \in \{rw, rv \text{ or } rm\}$ for $w,v \text{ and } m$ respectively, and $n$ is a number of total devices.

  \EndFor
  \EndProcedure
  \Procedure{\textcolor{blue}{deviceSelection}}{$w, v$}

  \For {$index$, device in sorted $D$}
        \If{$index < w$}
         \State   {deviceRole = worker}
         \ElsIf{$index >= w $ and $index <  (w + v)$}
         \State   {deviceRole = validator}
         \Else 
         \State deviceRole = miner
        \EndIf
    \EndFor
  \EndProcedure
  \end{algorithmic}
\end{algorithm}

\begin{table}[]
\caption{Notations used in Algorithm \ref{alg:role_mechanism}}
    \label{tab:notataions_role_mechanism}
    \centering
     \resizebox{0.7\columnwidth}{!}{%
    \begin{tabular}{|c|l|}
    \hline
      Notation  &  Meaning\\
      \hline
      $sv$  & Selection Value  \\
      $vrf$ & VRF output $\in [0,1]$ \\
      $s$ & Stake Value $\in [0,1]$\\
     $cp$ & Computation Power $\in [0,1]$\\
      $sh$ & Shape Value $\in [0,1]$ \\
      $wd$ & Wasserstein Distance $\in [0,1]$ \\
      $ls$ & Loss $\in [0,1]$ \\
     $\alpha_1$ & VRF factor \\
     $\alpha_2$ & Stake Factor\\
     $\alpha_3$ & Computation Factor\\
     $\alpha_5$  & Shape Factor \\
     $\alpha_6$  & Wasserstein Factor \\
     $\alpha_7$  & Loss Factor \\
     $w$ & Number of Workers \\
     $v$ & Number of Validators \\
     $m$ & Number of Miners \\
     $rw,rv,rm$ & Partition Ratio \\
       \hline
    \end{tabular}}
\end{table}

\subsubsection{Modules}
In the proposed device role selection mechanism, we calculate the selection value for a device based on 
a formula by considering different factors as, 
\begin{equation} \label{eqn:role_formula}
  \begin{aligned}
  sv = & \alpha_1 * vrf + \alpha_2 * s + \alpha_3 * cp  + \alpha_4 * sh \\
      & - \bigg\{ \alpha_5 * cd + \alpha_6 * wd + \alpha_7 * ls \bigg\}
  \end{aligned}
  \end{equation}
Those factors or modules of our proposed device role selection mechanism are explained here.

\begin{itemize}
    \item \textbf{Stake Accumulation ($s$)} :
    In blockchain-based FL, rewarding motivates clients to participate in the process.
    Thus, for implementation purposes, each role type is rewarded differently.
    For the worker, the reward will be based on the number of training epochs and the data size.
    Whereas validators get rewarded for the validation work of worker transactions.
    For miners, the task of mining blocks will contribute to its rewards.
    However, this strategy requires significant work and customization based on the needs of the system and the direction in which the work is intended to be carried out.
    For example, in a normal blockchain, miners are rewarded, not the normal nodes that create transactions.
    Whereas in FL, workers train the model, so they deserve more rewards than the server.
    However, every role (miner, validator, or worker) is important in blockchain-based FL.
    Therefore, developing a standard approach to measuring rewards for each role is quite complex and application-specific.
    
    \item \textbf{Computational Power ($cp$)} :
    In any network, actual computational power, i.e., the physical configuration of a device, can be different from what portion of its computation power is available.
    Even though the device might be computationally powerful initially, due to factors such as running other tasks simultaneously, computational power might vary in different time frames.
    So, to simulate this peculiar situation, the device's computational power is randomly changed in each communication round.
    We have computational power values ranging from 0 to 1 for implementation purposes.
    \item \textbf{No. of Devices and Roles} :
    We divide roles in a 5:2:1 ratio for implementation purposes.
    As shown in Figure \ref{fig:withselection}, the number of roles follows the ratio rule, with the number of a particular role changing in every round.
    However, without device selection, i.e., with random selection, 
    the number of different roles does not follow a particular pattern, as shown in Figure \ref{fig:withoutselection}.
    This might be an important aspect in terms of the manageability and predictability of the system.
    \begin{figure}[!h]
    \centering
    \begin{subfigure}[]{0.45\columnwidth}
    \includegraphics[width=\columnwidth]{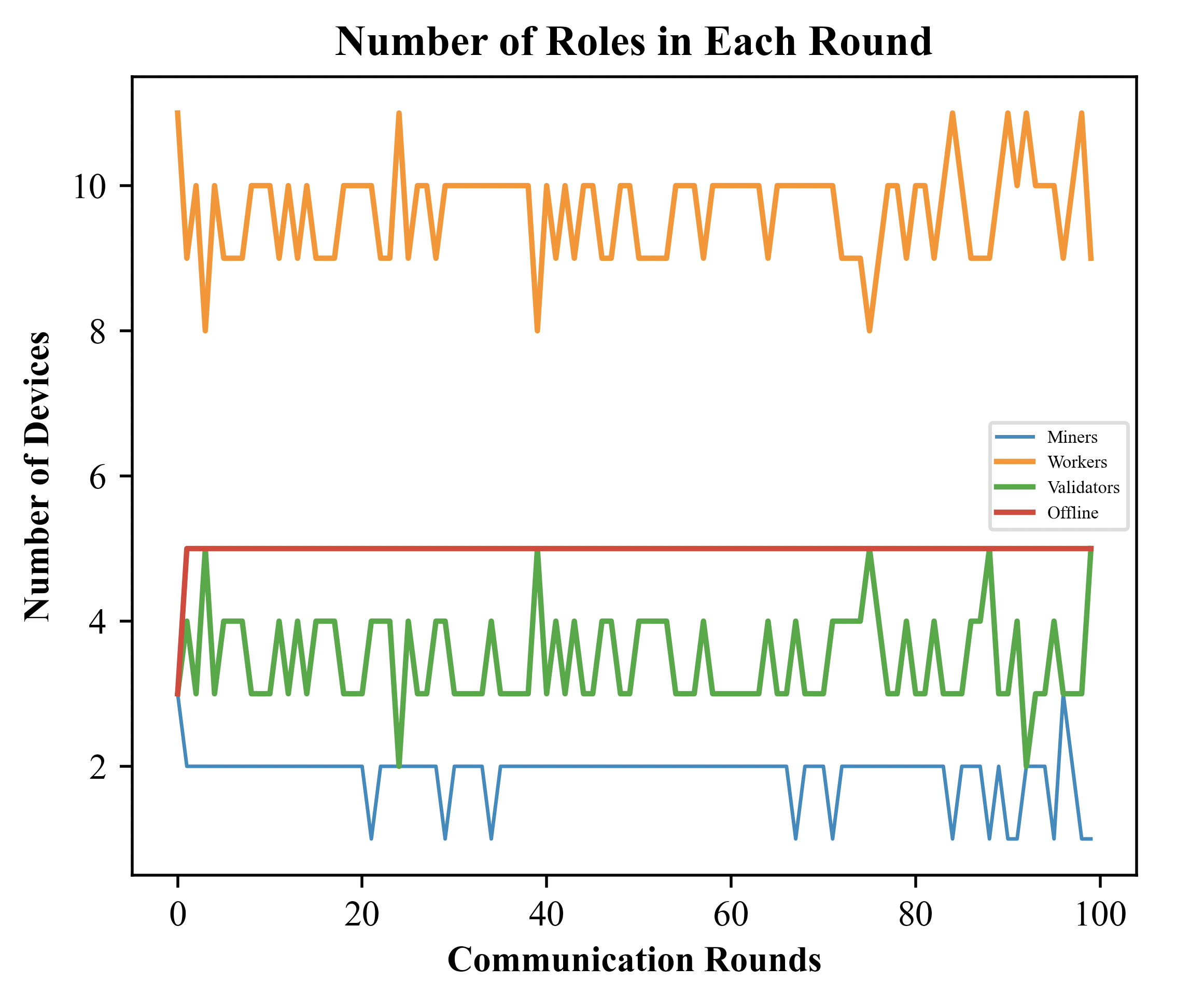}
    \caption{Fixed Ratio}
    \label{fig:withselection}
    \end{subfigure}
     \begin{subfigure}[]{0.45\columnwidth}
    \includegraphics[width=\columnwidth]{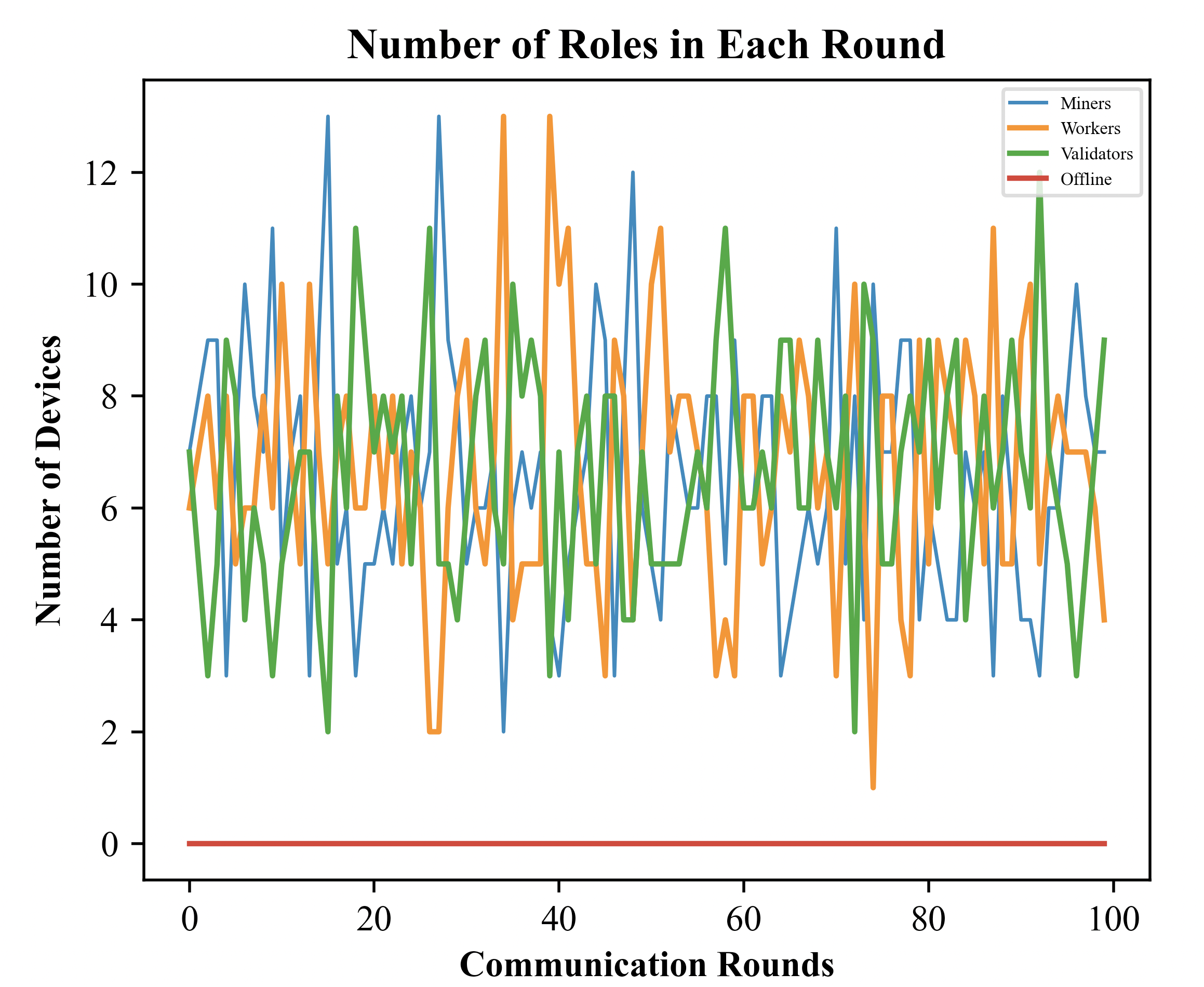}
    \caption{Without Ratio}
    \label{fig:withoutselection}
    \end{subfigure}
    \caption{No. of Roles in different communication rounds}
    \label{fig:numberofroles}
    
    \end{figure}
    \item \textbf{VRF for Randomness ($vrf$)} \label{module:vrf}:
    At the beginning of the communication round, decentralized VRF is used to add a randomness factor 
    to select the role for a device.
    That means that each device has a copy of the same VRF function.
    However, in the final round of selecting a winning miner to add the block to the blockchain, the output of the central VRF is compared with the initial output of the VRF of the device.
    To simulate, in our implementation, central VRF is a function that is separate from the device object.
   
    \item \textbf{Wasserstein Distance ($wd$)} :
    Wasserstein Distance is the difference between two data distributions \cite{huang2021shapley}.
    We have measured the difference in label distribution between each client's global dataset and sharded local dataset.
    The distribution of local data is an essential factor that affects model training \cite{data_distribution_Mustafa}.
    We have included this as one of the factors in deciding the selection value to determine a device role.
    Mathematically, if $l$ and $g$ are label distributions in the MNIST dataset of a local device and global dataset, respectively, then for a device $k$, the Wasserstein value will be
    $wd_k = g - l.$
    The value $wd_k$ is deducted from the selection value formula. 
   
    \item \textbf{Shape Value ($sh$)} :
    In this work, we have considered convergence contribution as shape value inspired by Shapely Concept.
    Each worker trains the global model parameters with their local data set to generate local models.
    Depending on how many epochs the worker performs and the dataset distribution, 
    the local models will perform differently with the test set data set of validators.
    Thus, against an average accuracy performance, each worker will have contributed differently in terms of learning convergence in each communication round.
    Therefore, on this basis, we consider this factor as a convergence contribution termed 'shape value'.
    To calculate the shape value, $sh$, we have considered the difference in the performance of trained models from a worker in the validator test data set and the average performance of all devices.
    If $w_k$ is a local model learned from a device $k$, $a_k$ is the accuracy in the validator test data set and $A$ is the average accuracy of all $K$ workers, then the shape value $sh$ for the device $k$ will be $sh_k = a_k - A$.
    
    \item \textbf{Cosine Distance ($cd$)}: Another factor considered for convergence is the Cosine distance between the two models. 
    The distance between the global and learned models is also calculated for each device and subtracted in equation \ref{eqn:role_formula}.
    
    \item \textbf{$\alpha$ parameters}: To calculate the selection value as described in Algorithm \ref{alg:role_mechanism}, 
    different parameters defined as $\alpha_1,\alpha_2, \alpha_3,
    \alpha_4,\alpha_5,\alpha_6, \alpha_7$ can be tuned depending on how dominant a particular factor we want it to be.
    Thus, if we only want to use a specific factor, then we can assign the values of all other parameters to zero.
\end{itemize}

\subsection{VRF implementation for Consensus}
\begin{figure}[!ht]
    \centering
    \includegraphics[width=\columnwidth]{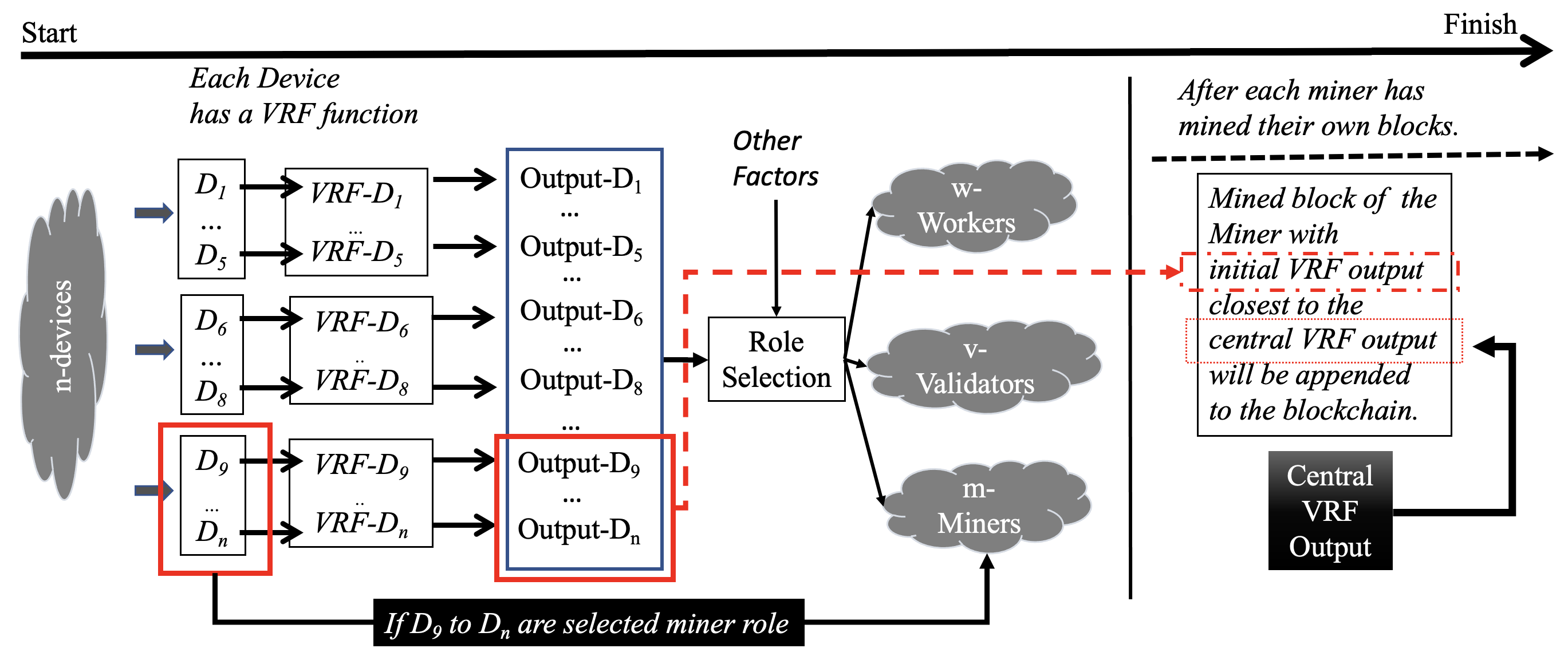}
    \caption{VRF Based winner block selection.}
    \label{fig:vrfconsensus}
\end{figure}

Along with the use of VRF for device selection (explained above in \ref{module:vrf}), its consensus design is as follows.
Considering a system where the miner is dominant, such as blockchain, our
proposed device selection mechanism selects miners
mainly with the highest stake. 
Thus, we can select any miner's block to be appended
to the blockchain.
However, to make it fair, we propose to employ VRF to randomly select the miner, as shown in Figure \ref{fig:vrfconsensus}.
In addition, we compare our initial VRF output that
we calculated in the beginning to select a role
for the device with another VRF output that is calculated
at the time of deciding the candidate block.
The only difference is that this will be computed only once
for all devices.
The miner with the initial VRF value closest to this
new VRF output will be selected as the winning miner.
However, workers are dominant if we consider FL and blockchain-based FL scenarios.
That means that miners' tasks will have less priority. 
Thus, even in that case, any miner can be selected using
VRF.
The algorithm is shown in \ref{alg:vrf} along with its notations in Table \ref{tab:notations_vrf_algorithm}.

\subsubsection{WOTS+ Verification for Distributed VRF} \label{definition:mvrf}
In our implementation of VRF, we have added WOTS+ signing of the VRF proof to produce a signature that
needs to be verified along with the normal verification of VRF. 
The steps involved in doing so are as below:
\begin{enumerate}
    \item Key generation
    \begin{itemize}
        \item Generate WOTS+ private key $wSk$ and public key $wPk$, $$(wSk,wPk) \leftarrow WOTS+\kgen(1^\lambda) $$
        \item Apply hash to $wSk$ as, $$wSkHash \leftarrow SHA(sSK)$$
        \item Obtain VRF private key $vSk$, $$vSk \leftarrow Encode(wSkHash) $$
        \item Employ VRF public key $vPk$ from private key as, $$vPk \leftarrow pubFunc(vSk). $$
    \end{itemize}
    \item Output computes, 
    $$(vOut, vProof, wSig) \leftarrow vrf(vSk, i/p, wSign(vProof). $$
    where, $vOut$ is output, $vProof$ is proof and $wSig$ is signature produced by signing
    $vProof$ using WOTS+ $wSign(vProof)$
    \item Verification consists of
    \begin{itemize}
        \item True or False $\leftarrow vVerify(vPk, vOut, i/p) $ 
        \item True of False $\leftarrow wVerify(vProof, wSig)$
    \end{itemize}
    where, $vVerify$ and $wVerify$ are verification functions and $i/p$ is the input.
\end{enumerate}

\begin{algorithm} [!htb]
  \caption{VRF for Selection Value and Miner Selection.} \label{alg:vrf}
  \begin{algorithmic}[1]
  \Procedure{\textcolor{blue}{forSelectionValue}}{$vrfOutput, v$}
  \State $vrfOutput \leftarrow dVRFunction()$ 
  \State $v \leftarrow ProbFunc(vrfOutput)$, where $v \in [0,1]$
  \State Use $v$ for selection value ($sv$) calculation.
  \EndProcedure
  \Procedure{\textcolor{blue} {minerSelection}}{$iV, fV$}
  \State Retrieve $iV$ for each $miner$, where $iV \in [0,1]$
  \State $cvrfOutput \leftarrow cVRFunction()$.
  \State $fV \leftarrow ProbFunc(cvrfOutput)$, where, $fV \in [0,1]$
  \For{$miner$ in $minerList$}
  \If{$min \mid miner_kiV - fV \mid$}
  \State $miner_k$'s block is the winner block.
  \EndIf
  \EndFor
  \EndProcedure
  \end{algorithmic}
\end{algorithm}

\begin{table}[!htb]
\caption{Notations used in Algorithm \ref{alg:vrf}}
    \label{tab:notations_vrf_algorithm}
    \centering
     \resizebox{0.8\columnwidth}{!}{%
    \begin{tabular}{|c|l|}
    \hline
      Notation  &  Meaning\\
      \hline
      $vrfOutput$  & Device VRF output   \\
      $cvrfOutput$ & Central VRF Output \\
      $dVRFunction()$ & Device VRF function \\
      $cVRFunction()$ & Central VRF function \\
      $iV$ & Initial VRF output $\in$ [0, 1] \\
      $v$ & Processed Device VRF output $\in$ [0, 1]\\
      $ProbFunc()$ & Converts bytes $\rightarrow$ [0,1] \\
      $fV$ & Processed central VRF Output\\
      $miner_k$ & Miner $k$ \\
      $miner_kiV$ & Device VRF output of Miner $k$ \\
      $minerList$ & List of all miners \\
      \hline
    \end{tabular}}
\end{table}

The consensus convergence can be quantified as follows:
$$T_{c} \sim Time(T_{vc} + T_{cvrf})$$
where $T_c$ is the time for consensus convergence, 
$T_{vc}$ is VRF output comparison, and $T_{cvrf}$ is the time taken for VRF output computation for
central VRF.
\footnote{
In our experiments, we observe that $T_{vc}$ is always smaller than the of Bitcoin ($\approx 10 mins$).}
\subsection{Some Questions in regards to proposed approaches}
A few questions regarding the approach and the choices made, and their advantages and disadvantages
that could be raised are answered below.
\begin{enumerate}[a.]
    \item \textbf{Why is Dilithium chosen?} Our consideration for a stateless signature can be any non-HBS PQC scheme.
    Thus, we could choose any stateless signature scheme for this purpose except for HBS schemes since they all have Merkle trees, state management, and authentication paths involved.
    Also, dilithium is faster than other PQC algorithms.
    It is also not an HBS scheme and, thus, does not involve any state management or authentication path computational complexities.
    That is the reason we did not choose SPHINCS+, which is also an HBS scheme. 
    SPHINCS+ is a stateless signature scheme but is still based on Merkle Tree Structure, which has its complexities of authentication path computation etc.
    
    \item \textbf{Why is stateful XMSS chosen? What is the problem with just using Dilithium only?} \\
    Dilithium is a signature scheme based on the hardness of lattice problems on module lattices.
    It is also one of the finalists in the NIST competition in the third round.
    Dilithium2 and Dilithium5 have signature sizes of 2420 bytes and 4595 bytes, respectively \cite{stebilaPostquantumKeyExchange2017}.
    Their public key sizes are 1312 bytes for dilithium2 and 2592 bytes for dilithium5.
    As we can see, that can be a problem in blockchain-based applications.
    So, signing a million transactions can easily lead to substantial storage requirements if we use only Dilithium.
    Above all, each blockchain network peer must download a complete copy of the entire blockchain.
    This can easily create a bottleneck in the performance of the network.
    For example, famous blockchains such as Ethereum have millions of transactions per day.
    Even with the best compression techniques, using massive schemes with big signature sizes thus will be costly in terms of space.
    Thus, we are using XMSS to sign the transactions, which has better signature and public-key sizes than other postquantum signature schemes when used with smaller tree heights.
    
    \item \textbf{What is the advantage of using a hybrid approach?} Currently, most PQC schemes used alone have key sizes and performance issues.
    Thus, with the hybrid approach, we can overcome some problems of both signature schemes.
    In terms of XMSS, it is stateful and based on the Merkle scheme.
    Only using XMSS has problems like the risk of state synchronization failures \cite{mcgrewStateManagementHashBased2016}, such as the same key being used to sign multiple messages or issues like running out of WOTS+ keys.
    Thus, whenever we generate a new XMSS tree, there is always a risk of being out of sync between critical updates and any network issues.
    Therefore, a hybrid approach seems to be a new way to look at how we can implement PQCs together.
    
    \item \textbf{What are some problems in the approach? Any other Challenges?} \\
    Even though XMSS is used with Dilithium, a new XMSS tree must be created once all OTS keys are used.
    In our approach, we recommend using smaller XMSS trees with a height of less than 10, giving $2^10 = 1024$ WOTS+ keys.
    This is because, with XMSS, one will have to generate all the keys initially.
    With a higher height, the key generation time will be long.
    Thus, we use small tree heights for faster key generation times with a trade-off of shorter single XMSS tree life, i.e. number of keys.
    We can do this because we can always generate a new tree in our approach.
    However, if longer XMSS tree life is of interest, i.e., more WOTS+ keys, 
    then bigger XMSS trees can be created, eventually costing some key generation time in the beginning.
    
    \item \textbf{Are there any assumptions in the approach?} Our hybrid approach is implemented and tested in a blockchain-based FL.
    Therefore, it is hard to say whether this approach is suitable or practical in different scenarios and with different hardware systems.
    
\end{enumerate}

\section{Experimental Results} \label{sec-experiment}

\begin{figure}[!ht]
    \centering
    \includegraphics[scale=0.08]{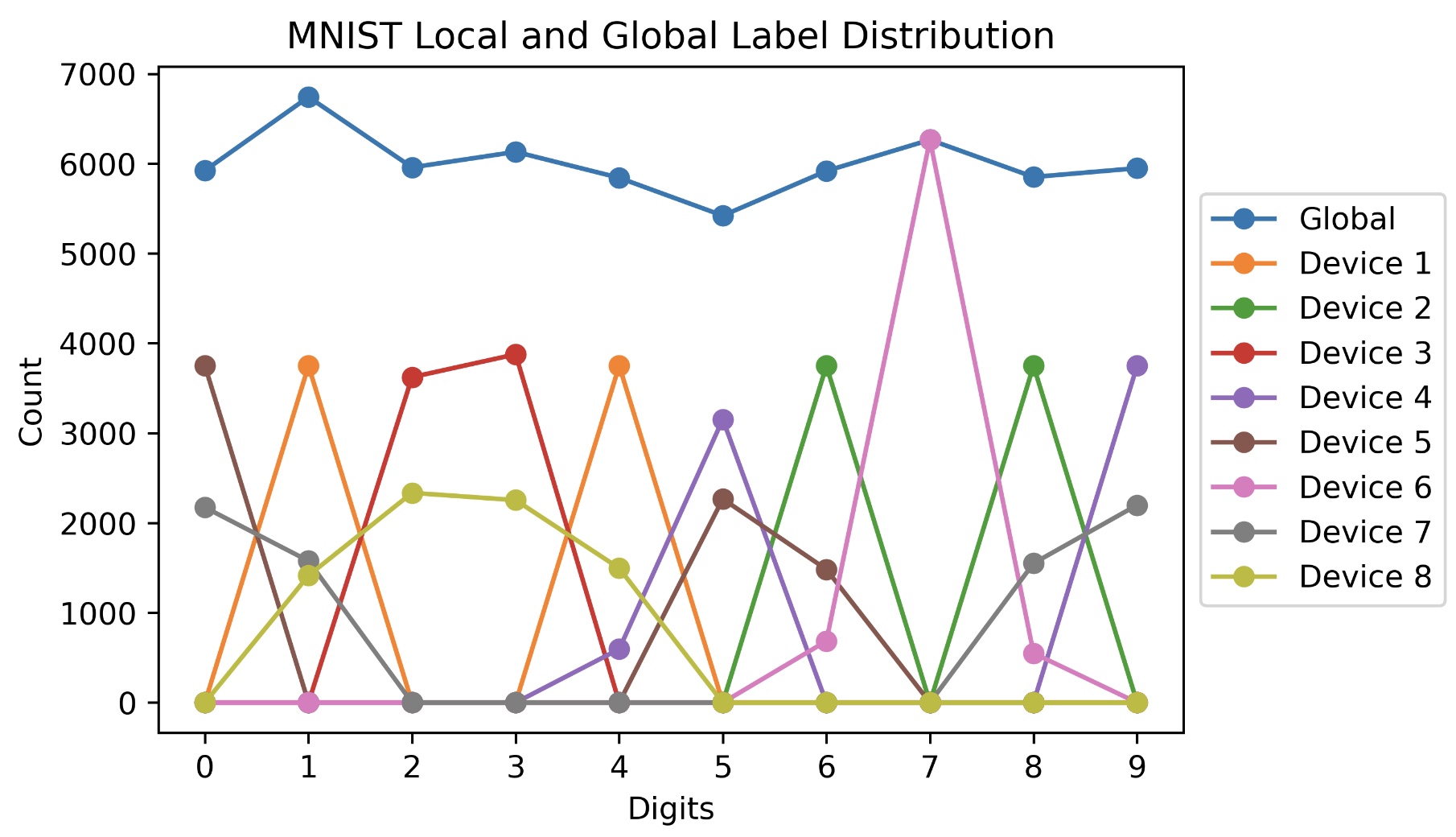}
    \caption{Distribution of Labels in each device over Dataset}
    \label{fig:label_distribution}
\end{figure}

To study the performance of the proposed hybrid signature approach,
device selection and VRF mechanism,  
we extended the implementation in BFL \cite{pokhrelFederatedLearningBlockchain2020a} and utilized open-source VBFL \cite{chenRobustBlockchainedFederated2021}
codebase with some required modifications per our needs.
The simulation is run on a local machine with a Quad-Core Intel Core i5 and on Google Colab Pro. 
The data set used is MNIST consisting of $70,000$ samples with each image size $28$ × $28$. 
For training, there are 60,000 samples, and the remaining $10,000$ are for testing purposes.
In circumstances where precision was not the main concern, small
sample sizes of less than $100$ were used to speed up the experimental process.
The \texttt{MNIST} data set is sharded at the beginning of the communication round
for all devices. 
The distribution of labels in each device training dataset and the global dataset is shown in Figure \ref{fig:label_distribution}. 
We can clearly observe that each device has a unique pattern in terms of distribution.
For example, in the global dataset, 
we can clearly find all labels (0 to 9) in contrast to a local device dataset.
Depending upon the need of the system, 
a specific device role can be prioritized 
to meet the intended goals.
There can be a few scenarios depending on that consideration.
\begin{enumerate}
    \item \textbf{Scenario 1 (Dominant Miners)} : 
    In blockchain settings, 
    miners are expected to have the highest computational power.
    They are also the devices that accumulate the most stakes throughout the process.
    From~\ref{eqn:role_formula}, we do not require a learning aspect.
    Thus, the values for $\alpha_4$, $\alpha_5$, $\alpha_6$ , $\alpha_7$ will be zero.
    The sorting of the list $D$ will be normal (in ascending order) to comparatively
assign the mining role to devices with higher computational power, higher stake and higher VRF output. 
    Then comes the validation, and eventually, the workers selected with the most negligible selection value.

    \item \textbf{Scenario 2 (Dominant Workers)} :
     In terms of blockchain-based FL or FL, 
     learning is the most important aspect. 
    Thus, workers are prioritized to be assigned to the devices with the best
    computational power, 
    most stake
    and contribution towards convergence.
    Thus, in algorithm \ref{alg:role_mechanism}, the list $D$ is sorted in reverse so that workers get better devices with higher computational power, higher stake, etc.
    \item \textbf{Scenario 3 (Random Selection)}: In this scenario, there is no priority in selecting a particular device based on any particular features.
    
    \item \textbf{Scenario 4 (Fixed or Random Ratio)}: It would also be interesting to see how the system behaves with a fixed ratio for several device roles and without a fixed ratio.
    This means that in one case there could be any number of workers, miners, or validators; in another, 
    we can have a fixed number of such device roles.
\end{enumerate}

\begin{figure}[!htb]
    \centering
    \includegraphics[width=\columnwidth]{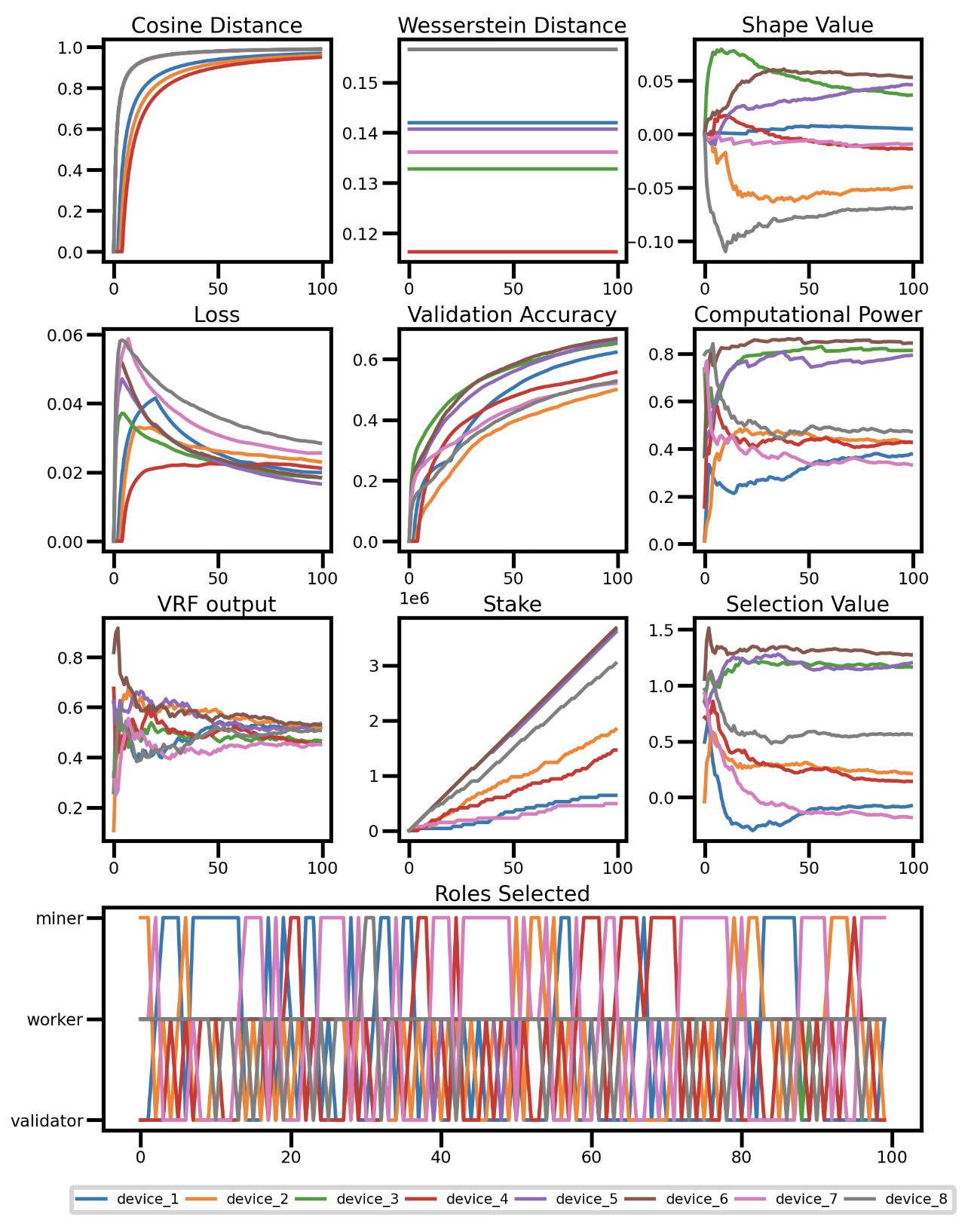}
    \caption{Device Behaviour in blockchain-based FL}
    \label{fig:device_behaviour_bfl}
\end{figure}

Based on the scenarios and discussions above, we first evaluate our fundamental presumption that \textit{devices
behave differently}.
In Figure \ref{fig:device_behaviour_bfl}, 
we can see how
a device behaves based on different factors 
(using \textbf{Scenario 2}) when the priority in device selection is given to the workers.
We can clearly see the difference in the Wasserstein distance ($wd$) in the
label distribution between each device dataset
against the global data set.
Device\_8 has the highest $wd$ value, whereas device\_4 has the smallest value.
For cosine distance $cd$, Device\_8 has the most significant value, whereas device\_4 has the smallest value.
We can clearly observe how closely $cd$, $wd$, $loss$, $sh$, and validation accuracy are related.
For example, with the highest value for $cd$ and $wd$, we can see that device\_8 has the most negligible value $sh$, that is, a negative contribution to the average accuracy.
Also, the loss value is the highest for device\_8 with poor validation accuracy.
In terms of computation power, device\_8 stands average among all devices, which is also reflected in the selected value, stake accumulation, and VRF output being average as well.
In the same Figure \ref{fig:device_behaviour_bfl}, we can also observe that  device\_4 is selected as a miner
only a few times, that is, it is selected mostly as a worker or validator as intended since in terms of data distribution, 
it has the minimum Wasserstein distance from the global
dataset label.
During validation, the loss value for the device\_4 is also average among the devices toward the end. 
This is also reflected in shape value, i.e., all workers' validation accuracy against average
accuracy.
For device\_4, for instance, its VRF value, computational power, and stake accumulation are all
average in value compared to other devices. 
Thus, we can conclude that higher cosine distance and Wasserstein distance lead to lower contribution towards validation accuracy, loss, and average accuracy (shape value).
All these factors contribute to the selected value.
It seems that device\_6, device\_3, and device\_5 are the best-performing devices
with higher values for shape value, validation accuracy, computational power, etc.
Thus, they are rarely selected as miners, as the proposed device role selection
mechanism is intended to do.
As more high-performance devices are selected as workers, the better the learning convergence
of the system.

\subsection{Analysis of Device Behavior}
\begin{figure}[!ht]
    \centering
    \begin{subfigure}[]{0.45\columnwidth}
    \centering
    \includegraphics[width=\columnwidth]{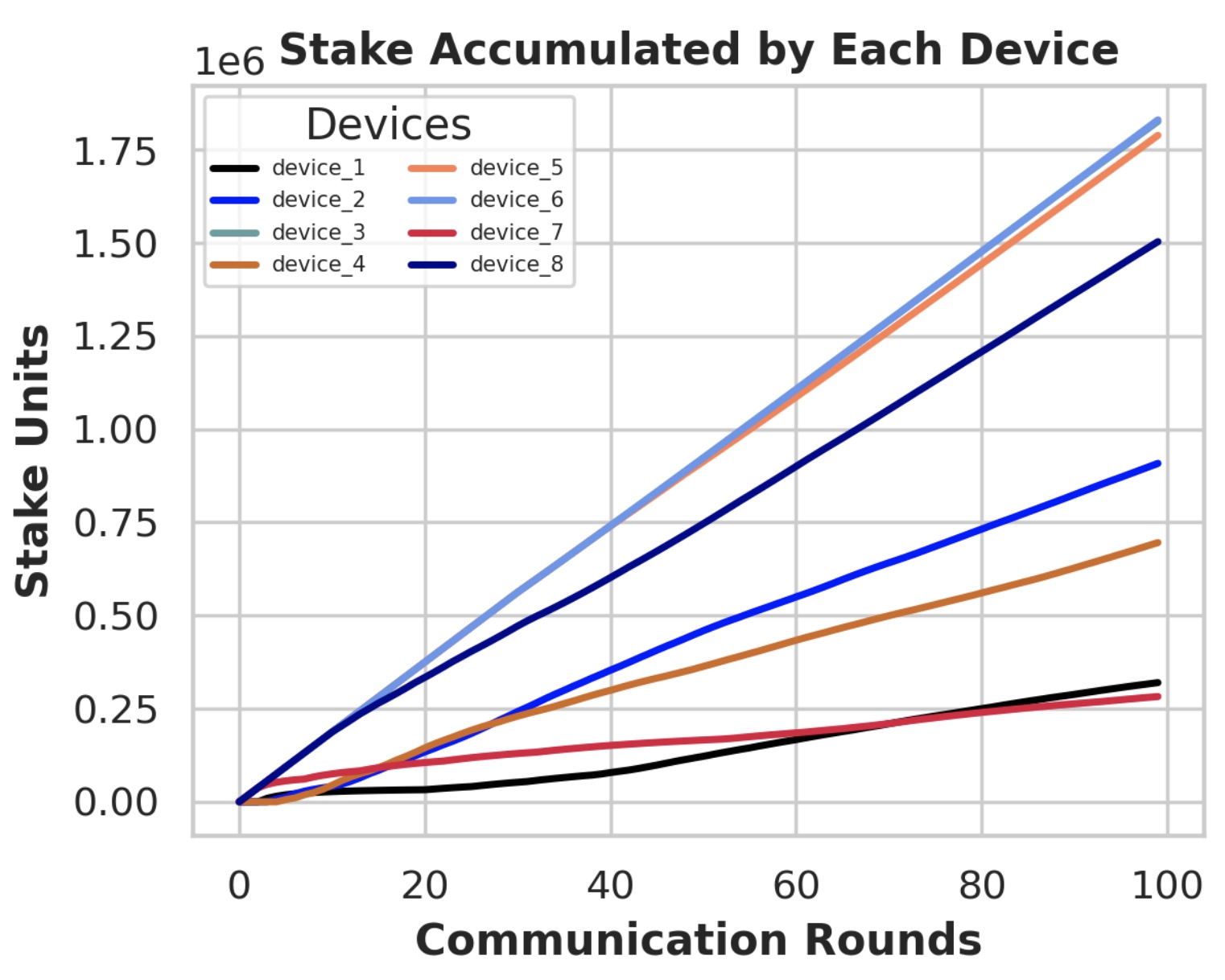}
    \caption{Stake Accumulation}
    \label{fig:stake_all}
    \end{subfigure}
    \centering
    \begin{subfigure}[]{0.45\columnwidth}
    \centering
    \includegraphics[width=\columnwidth]{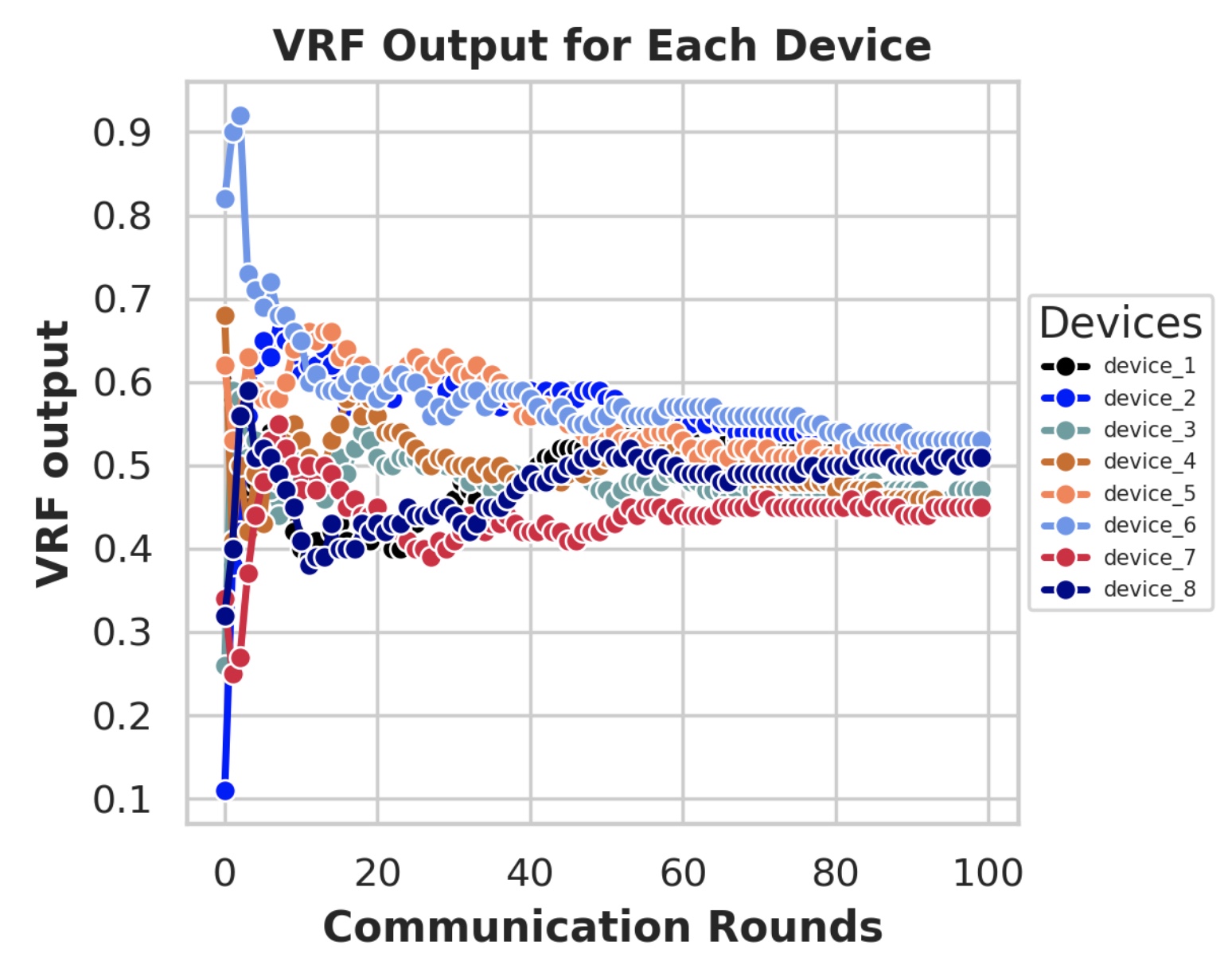}
    \caption{VRF Output}
    \label{fig:vrf_all}
    \end{subfigure}
     \centering
    \begin{subfigure}[]{0.45\columnwidth}
    \centering
    \includegraphics[width=\columnwidth]{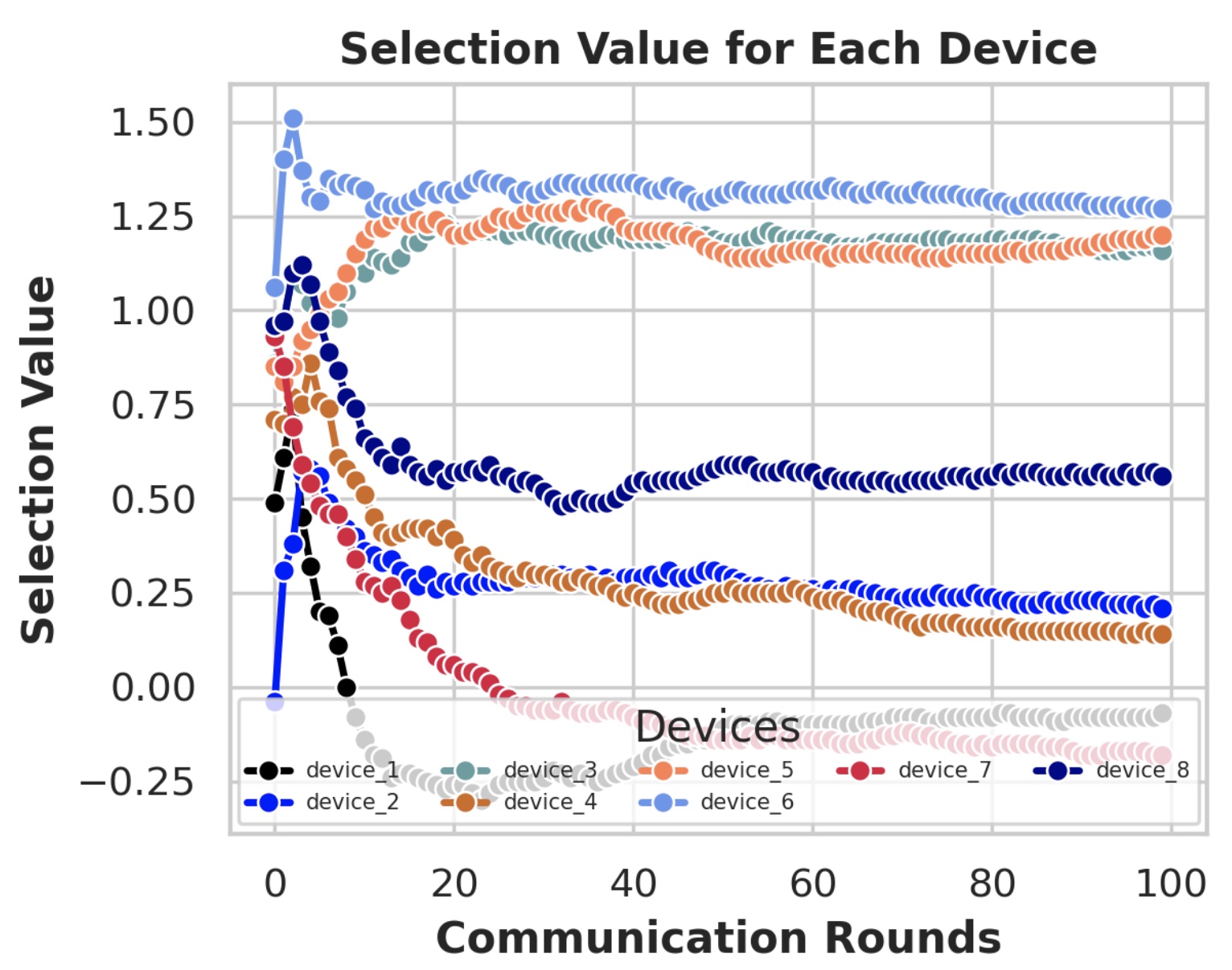}
    \caption{Selection Value}
    \label{fig:selection_value_all}
    \end{subfigure}
     \centering
     \begin{subfigure}[]{0.45\columnwidth}
     \centering
    \includegraphics[width=\columnwidth]{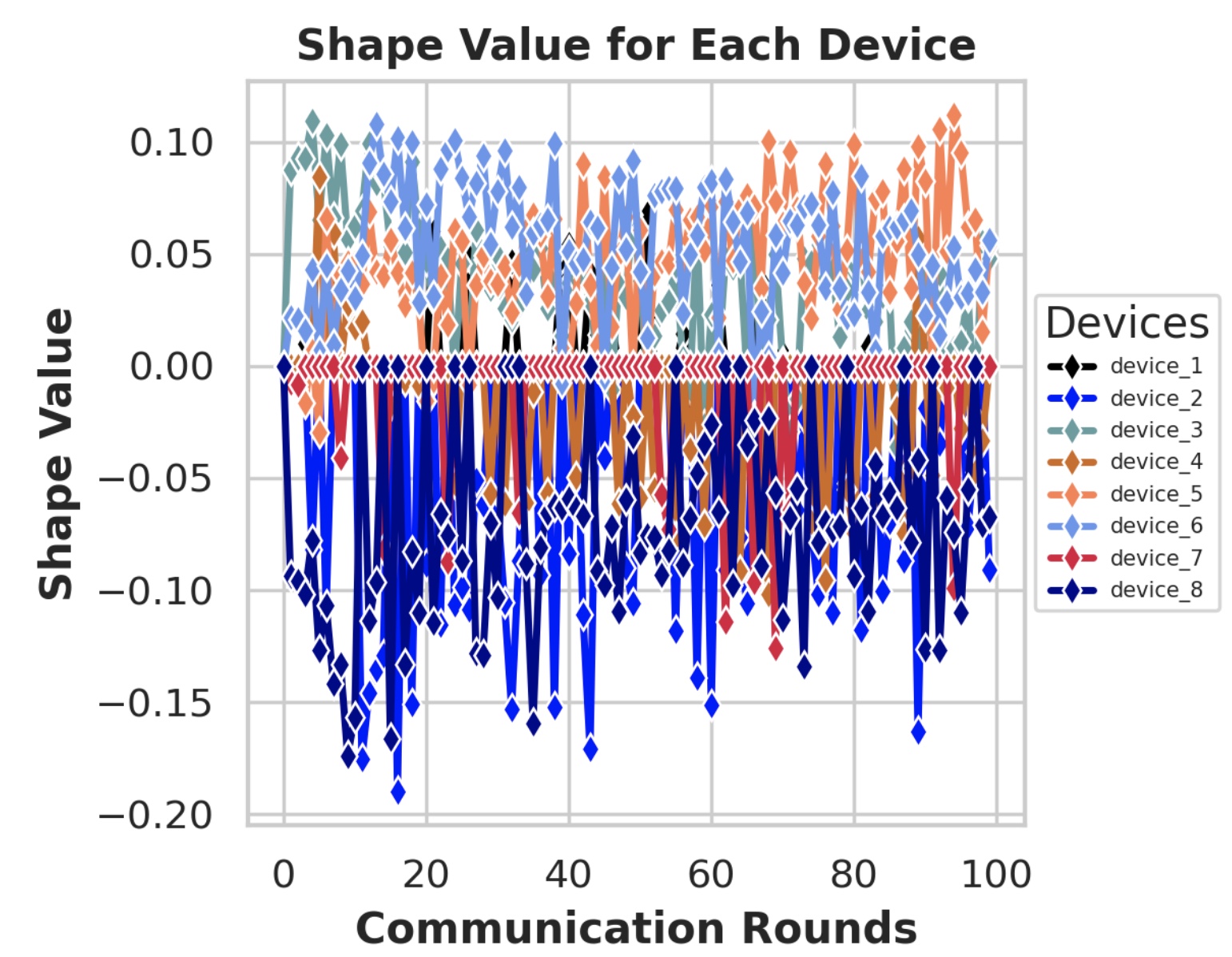}
    \caption{Shape Value}
    \label{fig:shape_value_all}
    \end{subfigure}
    \caption{Observation in blockchain-based FL}
    \label{fig:all_devices_bfl}
\end{figure}

To elaborate more, 
in Figure \ref{fig:all_devices_bfl}, we demonstrate our evaluation for the blockchain-based FL scenario. 
Figure \ref{fig:stake_all} shows that device\_6 has the highest stake accumulated throughout the $100$ communication rounds.
Whereas device\_7 and device\_1 struggle to accumulate more stakes.
In the mid-range, device\_2 and device\_5 accumulate average stakes.
Figure \ref{fig:vrf_all} shows the VRF values for each device.
The selection value and the VRF values
follow a similar pattern for most devices.
We can see a relative pattern among all these features due to the implemented device selection mechanism.
Selection values for devices 5, 6, and 3 are high because of their better contribution to average accuracy
and stake accumulation favored by by the VRF output value.
Thus, they will be selected workers more often, 
which eventually improves the performance of the system.
In figure \ref{fig:shape_value_all}, the device's contribution to
average accuracy is plotted.
From this figure, we can find that devices with a higher contribution towards accuracy have higher selection values and thus
are selected as workers to train the model further.

\begin{figure}[!htb]
    \centering
    \includegraphics[width=\columnwidth]{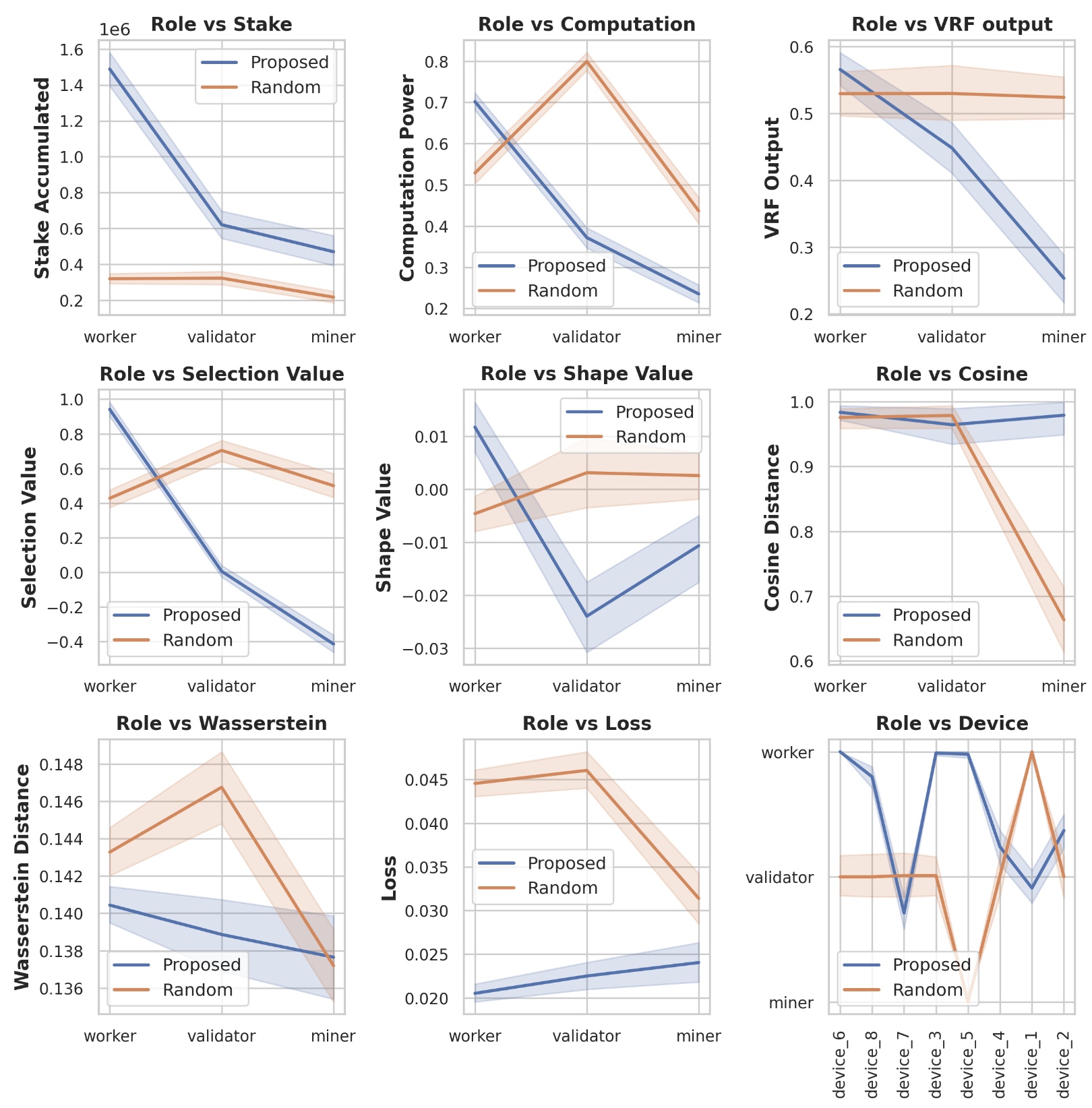}
    \caption{Role vs. other Factors in blockchain-based FL (Proposed vs. Random Selection)}
    \label{fig:rolevsothers_bfl}
\end{figure}

Figure \ref{fig:rolevsothers_bfl} demonstrates the impact of the proposed role selection mechanism (using \textbf{ Scenario 2}) against
random selection of roles (using \textbf{Scenario 3 \& 4}) in blockchain-based FL.
As shown in Figure \ref{fig:rolevsothers_bfl}, when following Scenario 2 (Dominant Workers), with the proposed approach,
we can see that workers are mainly selected with devices with higher stake value, more computational
power, VRF value, etc. 
In terms of contributing factors to convergence, such as shape value, workers
are selected for devices with a higher value than intended.
Which is almost the opposite or random when using random selection.
In terms of cosine distance between local learned parameters and global parameters and the
Wasserstein distance, there is a great difference between random selection
and our proposed selection approach.
With a random approach, devices with the lowest values for Wasserstein distance and the cosine distance
are more inclined toward minor roles.
This is not intended in the proposed selection.
Thus, the proposed selection is better and more balanced in that case, which would impact the
system less than the random approach.
Workers with the most negligible validation loss are assigned worker roles using the proposed approach
in contrast to the random approach.
If we see the devices against the roles plot, with the proposed selection, 
 devices 3, 5, and 6 are mostly selected worker roles.

\begin{figure}[!ht]
    \centering
    \includegraphics[width=\columnwidth]{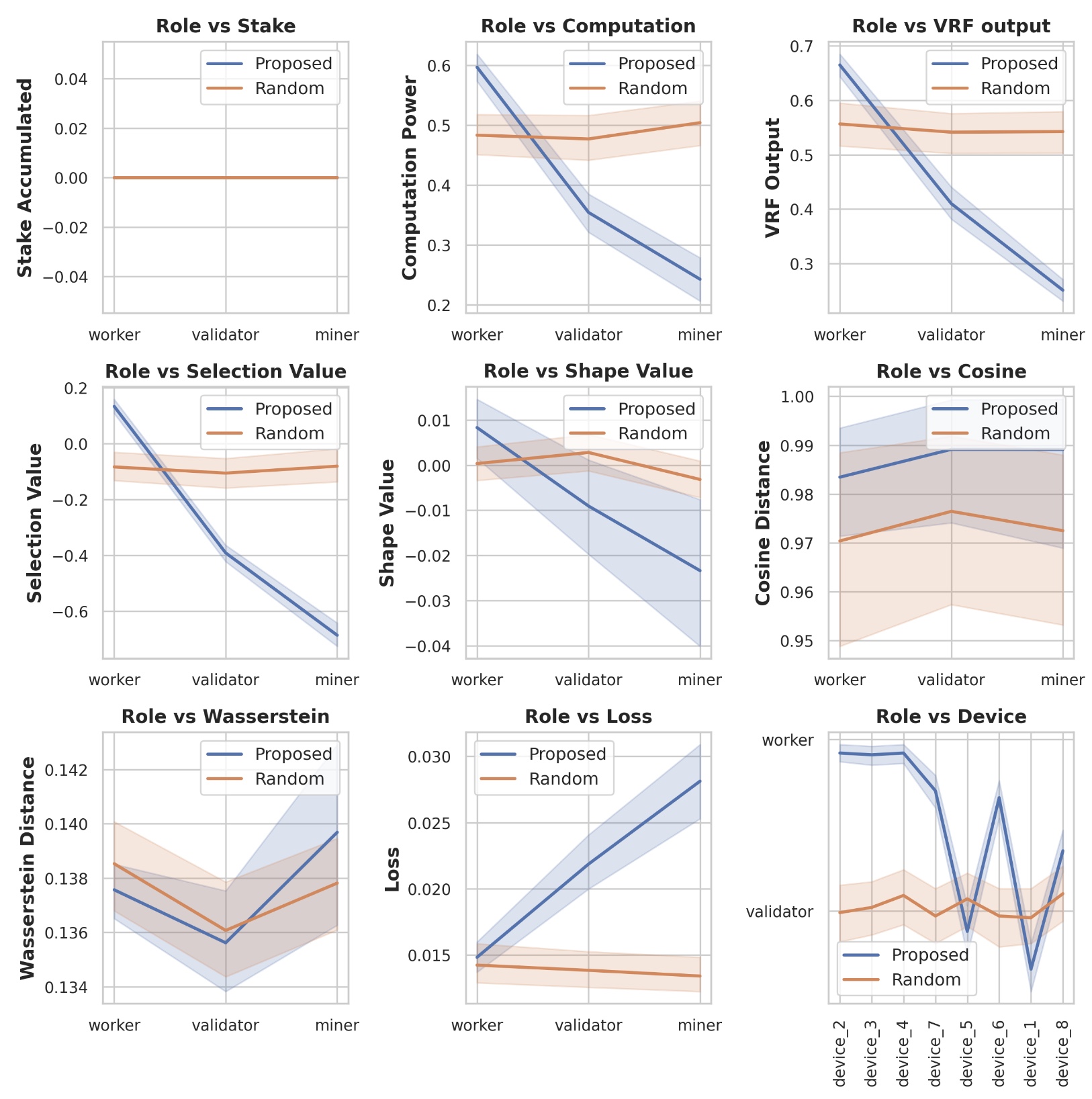}
    \caption{Role vs other Factors in FL (Proposed vs Random Selection)}
    \label{fig:rolevsothers_fl}
\end{figure}

A similar result is shown in Figure \ref{fig:rolevsothers_fl} for FL with a random selection mechanism and a proposed one.
For all devices, the stake is zero in FL.
The role validator only performs validation for the accuracy of the worker model, whereas the miner
is there in the plot only for reference purposes.
The computational power, VRF output, selection value, and shape value show similar patterns against the roles, and workers have higher values in the proposed method.
That is entirely different from the random approach.
In a way, for FL, with a random approach, the curve is almost flat for most results.
However, with the proposed method, devices assigned with the worker's role have lesser values for $cd$, $wd$, and the loss value.
That is how the selection mechanism is intended to work, accordingly.
This means that the workers chosen most of the time contribute more towards convergence and better
results in terms of learning.
This eventually improves the performance, as shown in Figure \ref{fig:randomvsproposed}, which
is explained in \ref{sec:accuracy_rvsprop}.

\subsection{Accuracy} \label{sec:accuracy_rvsprop}
\begin{figure}[!ht]
    \includegraphics[scale=.13]{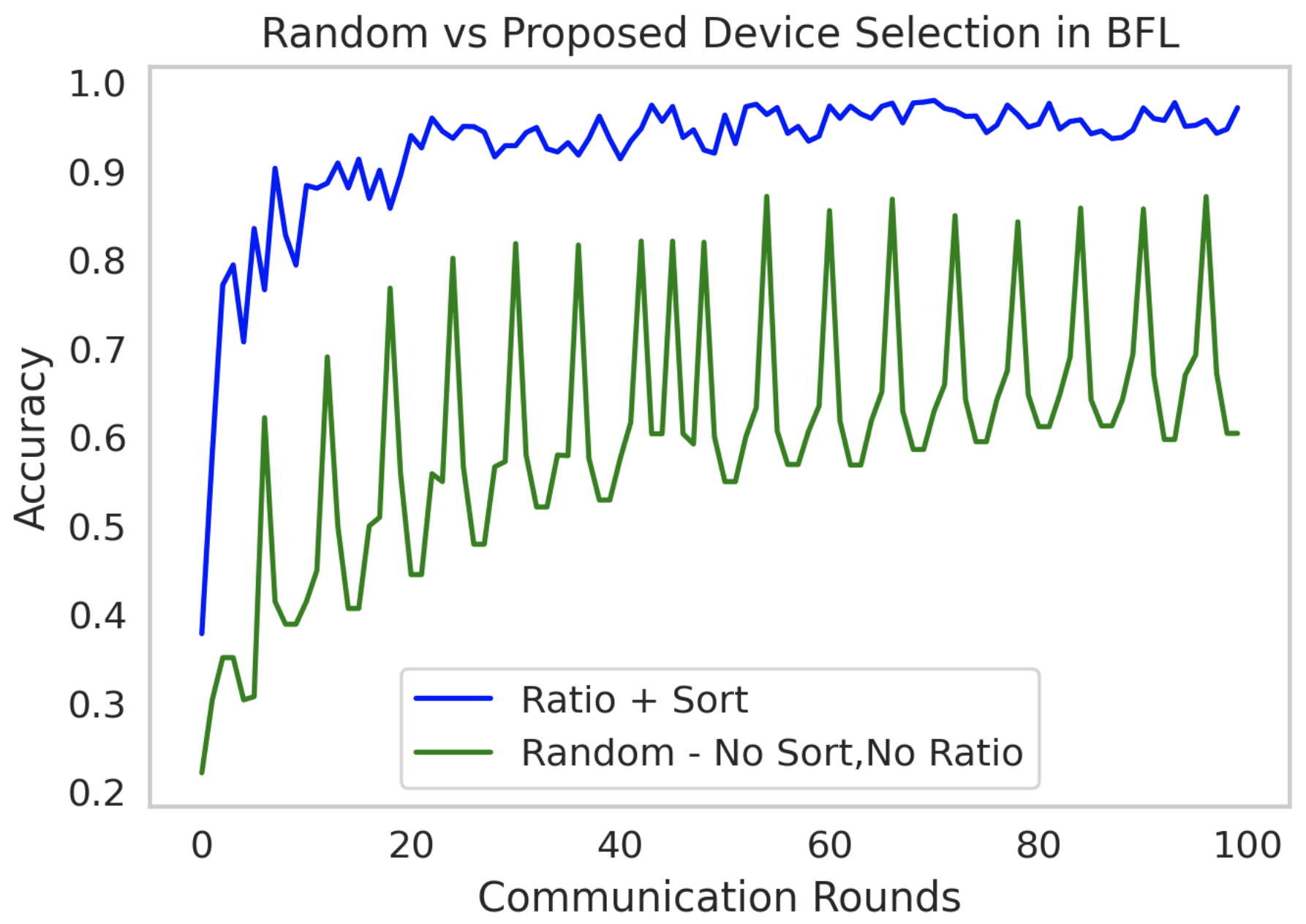}
    \includegraphics[scale=.14]{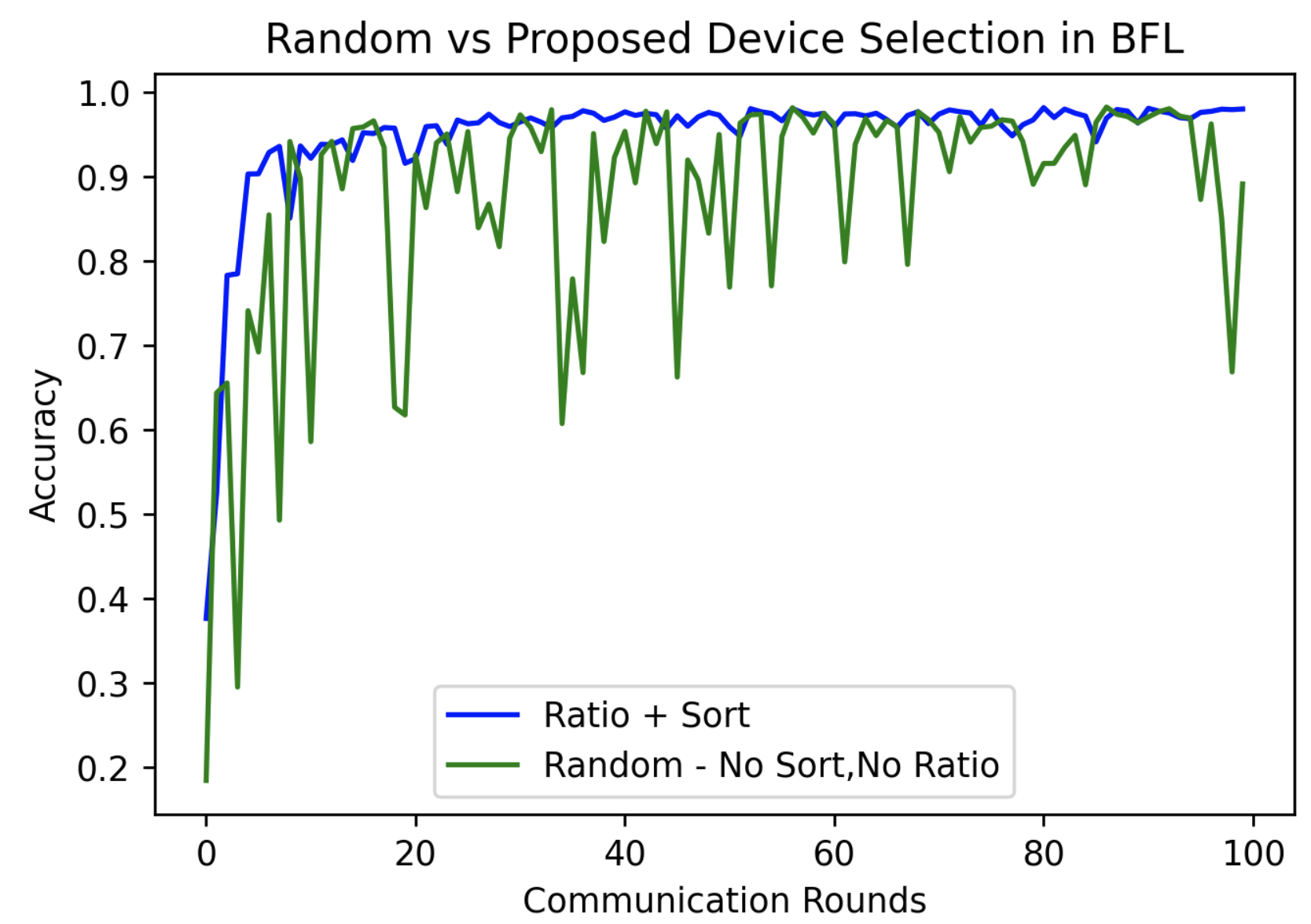}
    \caption{Comparison of the random vs. proposed device selection mechanism. Observe the difference in the evolution of accuracy with time over FL and blockchain-based FL systems.}
    \label{fig:randomvsproposed}
\end{figure}

In blockchain-based FL, running an experiment with eight devices with or without a ratio of 5: 2: 1 after sorting the devices based on the selected value is shown in Figure \ref{fig:randomvsproposed}.
It is interesting to see that
the ratio plays a vital role in
improving the performance of the system.
In contrast, a random approach of selecting devices with at least one device assigned to a particular role, 
there are high fluctuations in accuracy. 
We have similar observations in Figure \ref{fig:randomvsproposed} when
using both the proposed and random approaches.

\subsection{Communication Round Delay}
\begin{figure}[t]
    \centering
    \includegraphics[scale=0.3]{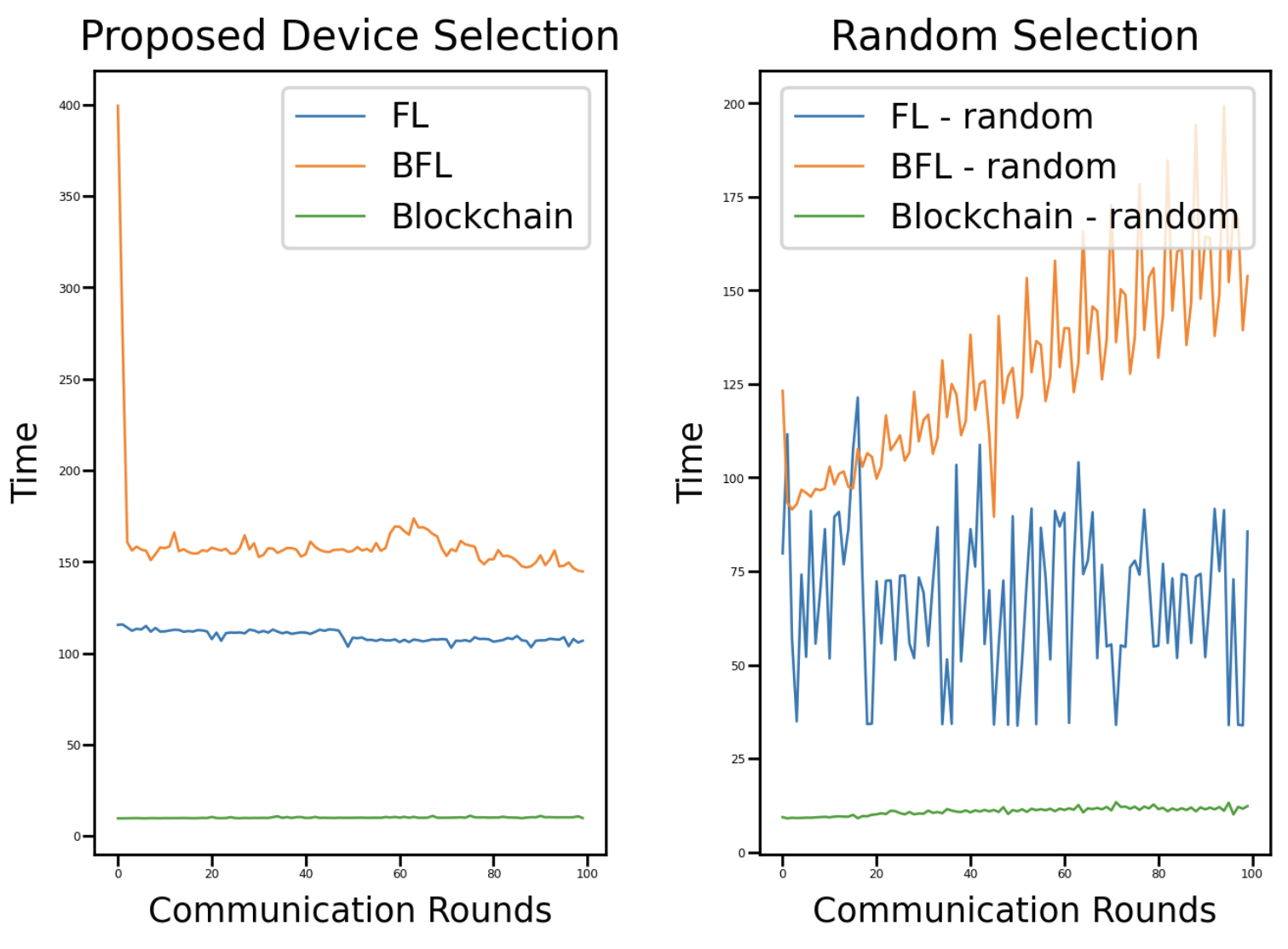}
    \caption{Time to complete each communication round}
    \label{fig:comm_time_all}
\end{figure}

Simulation results of communication round delay
for three networks blockchain, blockchain-based FL, and FL
are shown in Figure \ref{fig:comm_time_all}.
The results show that the
blockchain takes the least time to complete its communication rounds.
But it should be considered that the winner-miner block selection in
the implementation is made by comparing VRF. 
Thus, depending on the type of blockchain network and the consensus
used,
the communication round delay might be greater than the results shown here.
However, in comparison between FL and blockchain-based FL, blockchain-based FL is slower than FL.
In contrast to only the training time involved, as in FL, 
blockchain-based FL also has tasks associated with miners and validators, adding up to more
time for the completion of communication rounds.
The effect on the communication time of the proposed device selection is
shown in the right plot of the same figure \ref{fig:comm_time_all}.
It is also clear that in the random approach, 
there is a lot of fluctuation in time to complete a communication round.

\section{Experimental Results of PQC vs CLassical Cryptography}

In this section, the experimental analysis of the performance of different PQC schemes such as Dilithium, Falcon-1024, and SPHINCS + -SHA256-256s-robust is compared to RSA-1024.
These experiments were done on a local computer using "liboqs" and
"liboqs-python" libraries for Dilithium, Falcon, and SPHINCS+.
For XMSS, the "winternitz" library is used to generate WOTS+ keys and their performance.
Whereas for RSA signature schemes, the \textit{ PyCryptodome} library is used.
\textit{It is important to note that as experiments are performed using different available libraries, the data structure used to store critical hashes might impact the results shown here}. This might need further investigation.

\subsection{Cryptography Comparison}
\begin{figure}[!h]
    \centering
    \begin{subfigure}[]{0.345\columnwidth}
    \centering
    \includegraphics[width=\columnwidth]{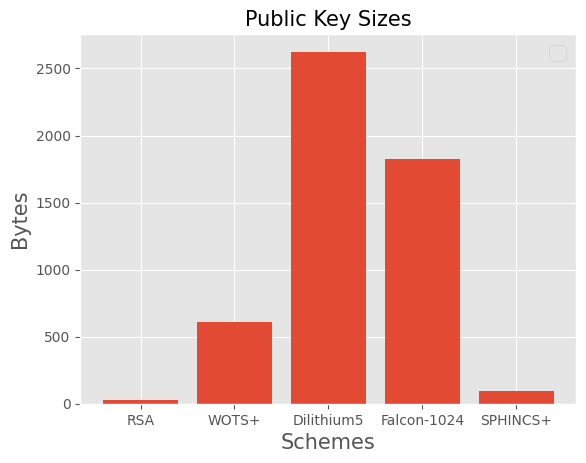}
    \caption{Public keys}
    \label{fig:public_keys_comparison}
    \end{subfigure}
    \centering
    \begin{subfigure}[]{0.345\columnwidth}
    \centering
    \includegraphics[width=\columnwidth]{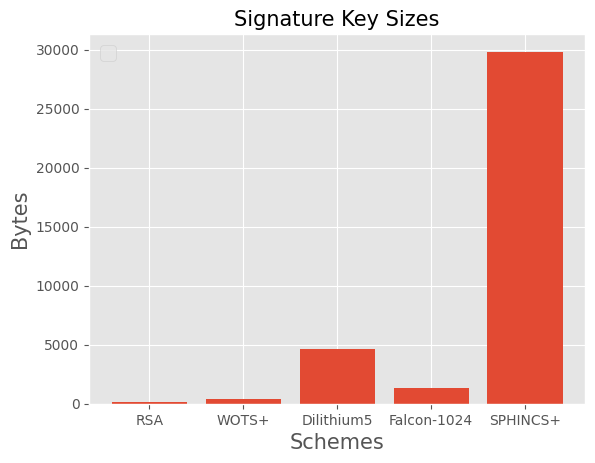}
    \caption{Signature Sizes}
    \label{fig:signature_keys_comparison}
    \end{subfigure}
     \centering
    \begin{subfigure}[]{0.345\columnwidth}
    \centering
    \includegraphics[width=\columnwidth]{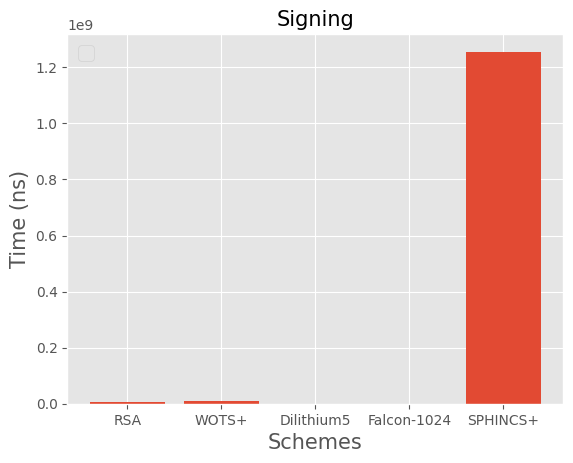}
    \caption{Signing Time}
    \label{fig:Signing Comparison}
    \end{subfigure}
     \centering
     \begin{subfigure}[]{0.345\columnwidth}
     \centering
    \includegraphics[width=\columnwidth]{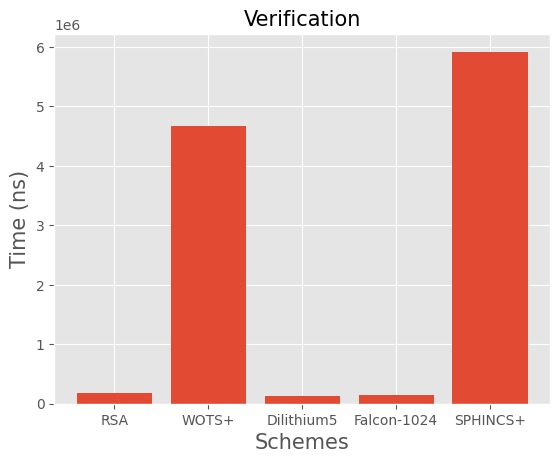}
    \caption{Verification Time}
    \label{fig:verifications}
    \end{subfigure}
    \caption{Cryptography Comparisons}
    \label{fig:cryptography_comparisons}
\end{figure}

In Figure \ref{fig:public_keys_comparison}, Dilithium and Falcon have much larger key sizes in terms of public keys. It seems the WOTS+ public key is also a bit bigger than RSA and SPHINCS+. 
It is clear that the RSA signature scheme is better in both key sizes, as shown in Figures
\ref{fig:public_keys_comparison} and \ref{fig:signature_keys_comparison}.
This is the reason, we propose the use of the XMSS signature scheme for signing transactions, which is the hash-based scheme that has smaller public key sizes than that of Dilithium and Falcon.
In terms of verification, Dilithium and Falcon are comparably better than WOTS+ and SPHINCS+ and are almost equal to RSA.
While, for signing, SPHINCS+ is very slow.
From the results, we can conclude that Dilithium and Falcons shine in their performance of key generation and signing, whereas XMSS shines with its small public key size.
Thus, our work is wholly focused on the idea that using two PQC schemes, such as XMSS and Dilithium (or Falcon), might be complementary and beneficial in many circumstances
depending upon the need of the application.

\begin{figure}
    \centering
    \includegraphics[scale=0.34]{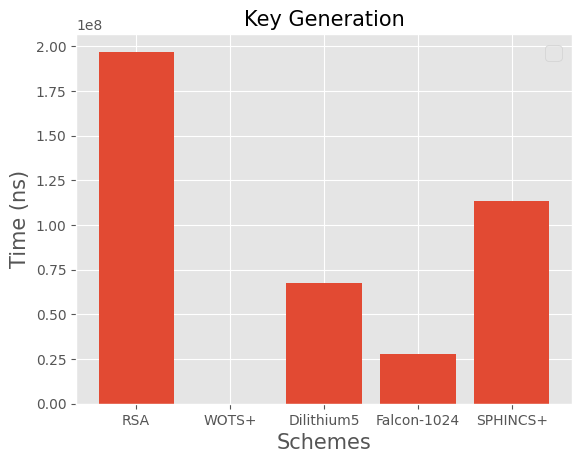}
    \caption{Comparision of Key Generation time}
    \label{fig:key_generations}
\end{figure}
From Figure \ref{fig:key_generations}, in terms of key generation,
RSA seems slower than other signature schemes. 
Here, we can clearly see that WOTS+ signature schemes are faster than other PQC schemes.
This is one of the best advantages of using XMSS signature schemes as it uses WOTS+ keys.

For a typical example as ours, i.e. Post Quantum Secure Blockchain-Based application, 
we proposed the idea of integrating schemes like XMSS and Dilithium (or Falcon).
As mentioned prior as well, the notion of using these schemes in a hybrid manner can be a viable and novel approach to secure our digital world against quantum threats in the best possible manner. 

\subsubsection{Key and Signature Sizes}
As shown in \ref{fig:public_keys_comparison} and \ref{fig:signature_keys_comparison}, all PQC schemes such as
Dilithium, Falcon, and SPHINCS+ are considerably larger than WOTS+ and RSA.
It is important to note that the Python programming framework $sys.getsizeof()$ is used for this purpose. 
This is to clarify that XMSS and other PQC schemes have a bit different implementation for storing its hash values for keys, i.e. Python data types.
This makes our choice of XMSS for signing the transactions
more practical than other PQC schemes.
However, in terms of performance, dilithium and falcon are
also quite good in comparison to WOTS+ and RSA.
SPHINCS+, however, is slower than others.
Even in terms of signature size, XMSS is more practical. 
Dilithium has a larger signature size.
Even though the performance of Dilithium is quite good, its key and signature sizes are a huge problem if implemented in a
blockchain network which keeps growing in size.
With thousands of transactions to be signed, verified, and validated, many signatures can cause storage problems.
Thus, our approach does not use dilithium for signing transactions. 

\subsection{Performance Analysis of PQC over BFL}

\subsubsection{Transmission Delay and Transaction Size}
\begin{figure}
    \centering
    \begin{subfigure}[]{0.4\columnwidth}
    \includegraphics[width=\columnwidth]{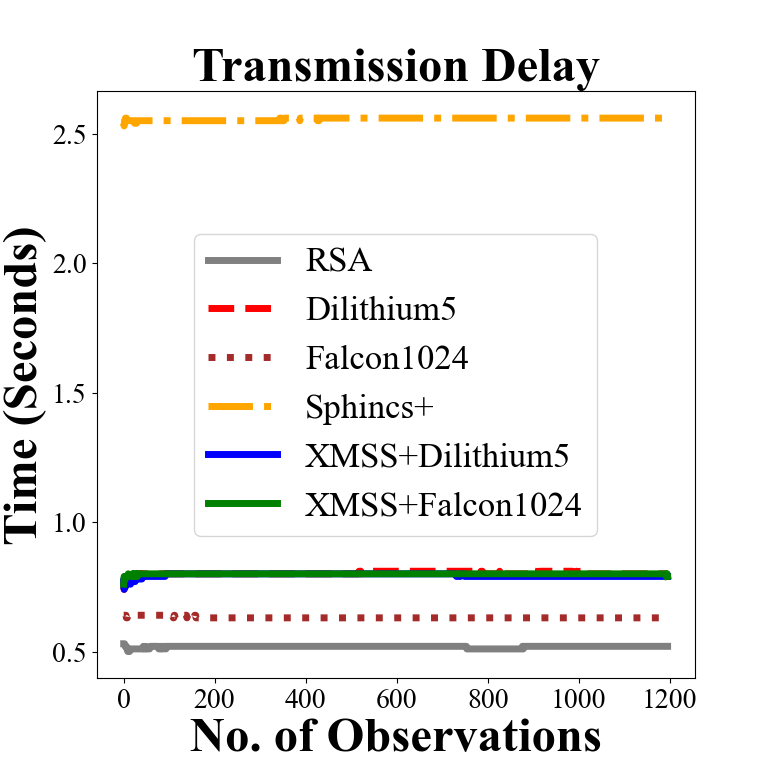}
    \caption{Transmission Delay}
    \label{fig:transmission_delay}
    \end{subfigure}
    \begin{subfigure}[]{0.45\columnwidth}
    \includegraphics[width=\columnwidth]{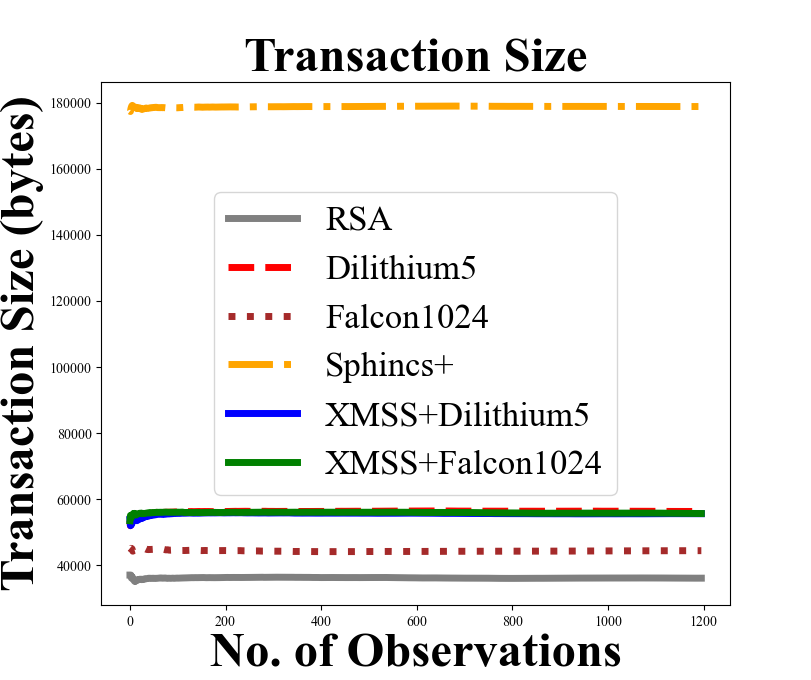}
    \caption{Transaction Size}
    \label{fig:transaction_size}
    \end{subfigure}
    \caption{Transmission Delay and Transaction Sizes}
    \label{fig:delayandsize}
\end{figure}

The figure \ref{fig:transmission_delay}  shows how using different
signature schemes impact the performance of the system.
Here, we compare transmission delay, which refers to the data transfer
per second.
SPHINCS+ is the slowest among all.
Whereas, Dilithium, XMSS with Dilithium, and XMSS with Falcon all fall in the same area on the graph.
However, Falcon seems to be better with RSA being the best in terms of
transmission delay.
The pattern is similar to the transaction sizes shown in \ref{fig:transaction_size} as well.
With hybrid implementation, only the XMSS signature and public key
are included in the transactions.

\subsubsection{Block Generation and Communication Round Time}
\begin{figure}[!tbp]
    \centering
    \begin{subfigure}[]{0.44\columnwidth}
    \centering
    \includegraphics[width=\columnwidth]{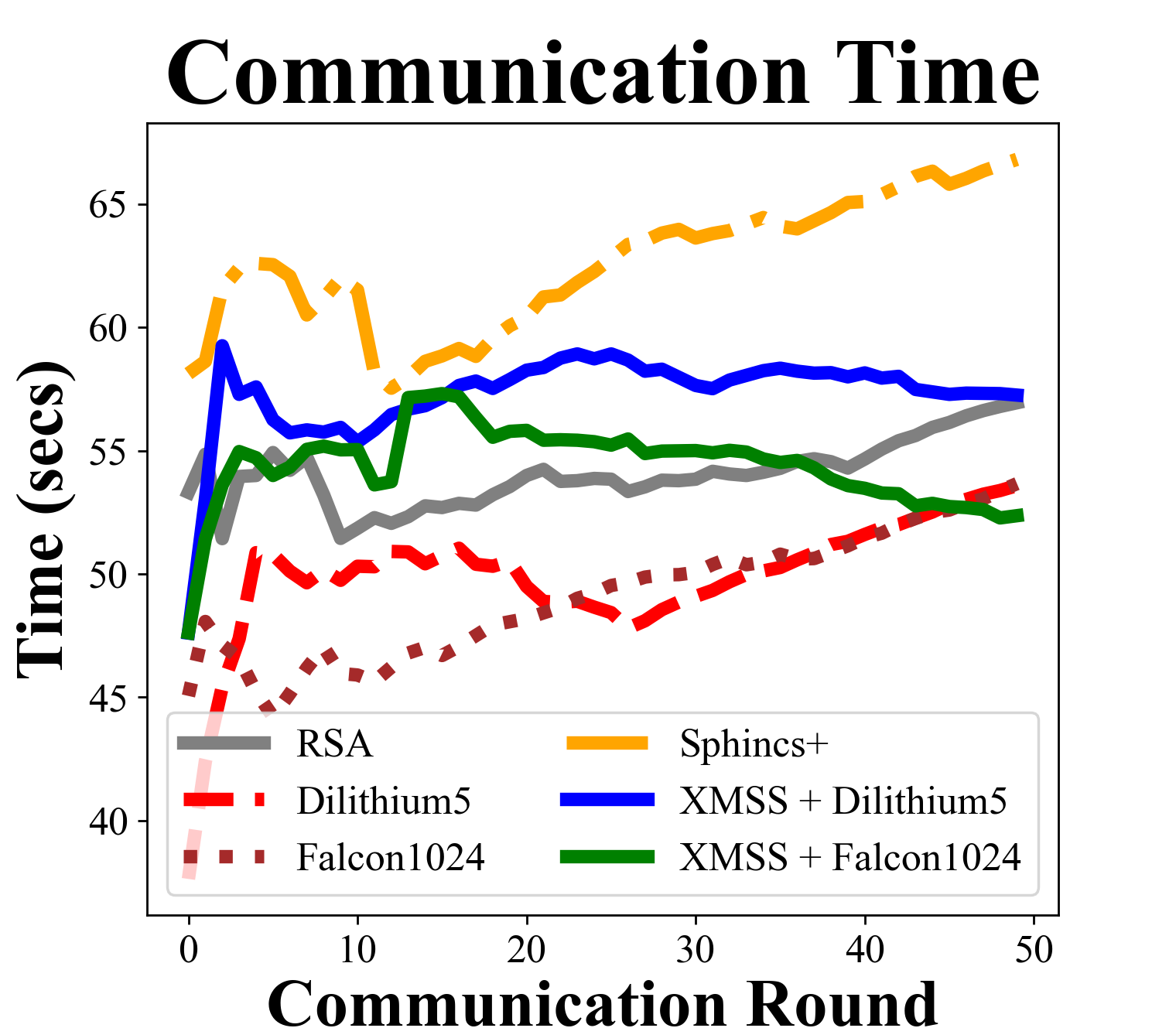}
    \caption{Communication Time}
    \label{fig:communication_round}
    \end{subfigure}
    \begin{subfigure}[]{0.45\columnwidth}
    \centering
    \includegraphics[width=\columnwidth]{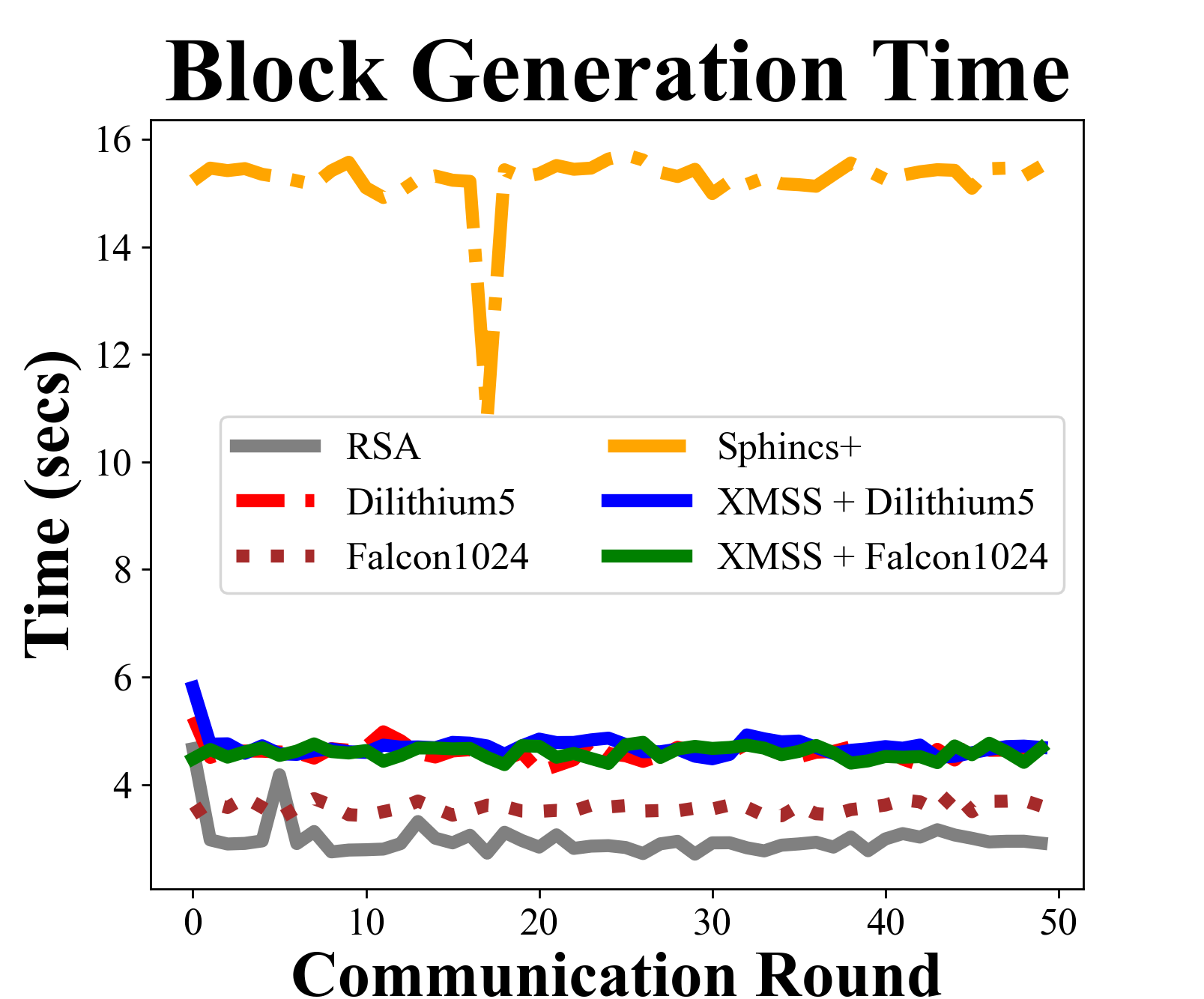}
    \caption{Block Time}
    \label{fig:block_generation}
    \end{subfigure}
    \caption{Block Generation and Communication Round Time}
    \label{fig:communicationandblock}
\end{figure}

The communication overhead of the system due to the implementation of PQC is shown in Figure \ref{fig:communicationandblock}.
The performance of the cryptography used directly affects the block generation time in the blockchain network.
With slower key generation, signing, and verification time, there is also a delay in the whole system, leading to slow consensus.
In Figure \ref{fig:block_generation}, 
Sphincs+ with its block generation time is much slower
than other PQC schemes. 
However, RSA is the fastest, and Falcon next to it performs better than other schemes.
With the hybrid approach, its performance is similar to that of the dilithium scheme.
In terms of the total time to complete each communication round, as shown in Figure \ref{fig:communication_round}, 
SPHINCS+ is still the slowest, while Dilithium and Faclon perform better.
The hybrid approach also needs improvement to catch up with the RSA scheme.
However, the performance of the hybrid approach is quite comparable with Dilithium or Falcon, or even RSA considering that it uses two signature schemes.

\subsubsection{Accuracy}
\begin{figure}
    \centering
    \begin{subfigure}[]{.42\columnwidth}
    \centering
    \includegraphics[width=\columnwidth]{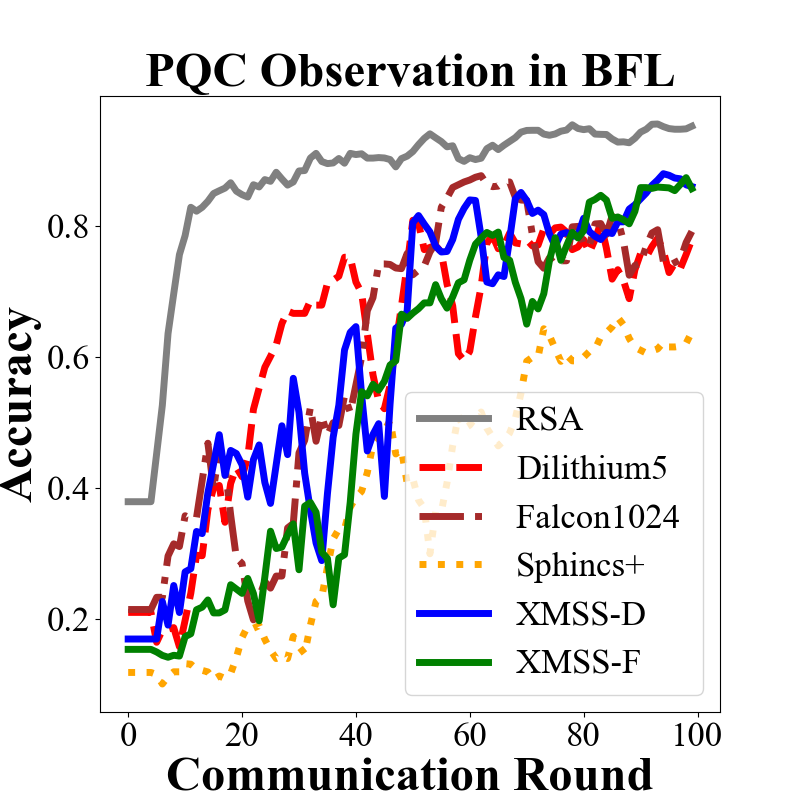}
    \caption{Only replacing PQC}
    \label{fig:onlypqc}
    \end{subfigure}
    \centering
    \begin{subfigure}[]{.45\columnwidth}
    \includegraphics[width=\columnwidth]{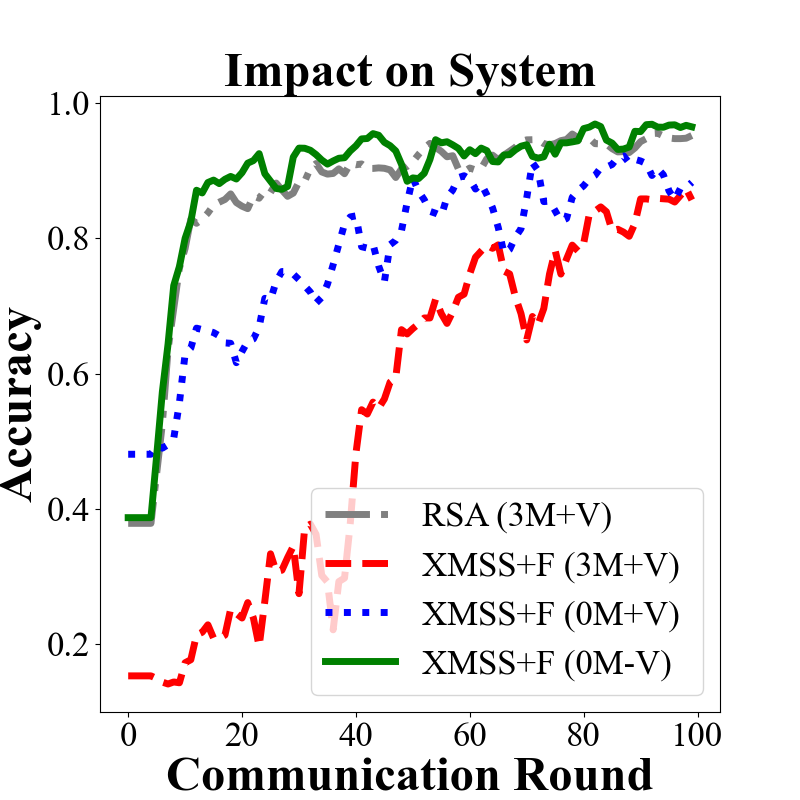}
    \caption{With or without validation}
    \label{fig:malicious}
    \end{subfigure}
    \caption{Accuracy observations with PQC replacement}
    \label{fig:accuracyonlycryptography}
    
\end{figure}

With the initial assumption that the cryptography used should not affect the accuracy
of the BFL system, it was interesting to see that just
replacing RSA with other PQC schemes in the original VBFL code base \cite{chenReportPostQuantumCryptography2016}
(without changing anything else), the accuracy achieved was not like
when RSA was used.
The results are shown in figure \ref{fig:onlypqc} which are obtained after running the simulation,
with exactly the same parameter values presented in their paper \cite{chenRobustBlockchainedFederated2021}, except for the validation threshold where Dilithium5, Falcon-1024, RSA, and XMSS used a bit different value than XMSS-D and XMSS-F.
All simulations were done with 3 malicious nodes and, altogether, 20 devices.
After doing some more experiments, we came to realize that the main reason would be that
the validation mechanism is impacted and does not perform as needed due to communication overhead.
Thus, to further verify, we replaced RSA with the XMSS+F hybrid scheme and ran the simulation with RSA (3M + V), XMSS + F (3M + V), XMSS + F (0M+V), and finally XMSS + F (OM-V: no malicious nodes and no validation mechanism, but assuming the presence of a perfectly functioning malicious node detection system).
Here, M refers to malicious nodes (3M means 3 malicious nodes, etc.) and "+/- V" refers to with or without the validation mechanism implemented in \cite{chenRobustBlockchainedFederated2021}.
As shown in Figure \ref{fig:malicious}, it is clear that PQC in itself cannot and does not impact federated learning in itself.
The green curve line using a hybrid signature scheme is the result when there is no malicious node and there is
no validation mechanism as well.
In other words, for the simulation, we have assumed the presence of a perfect validation mechanism
that is always correct for detecting malicious nodes.
This implies that PQC implementations directly impact system performance.

\subsubsection{Comparison between different PQC schemes}
In this section, we show some experimental results of the implementation of different PQC schemes in the BQFL system. 
The PQC schemes compared are Dilithium2, Dilithium5, Rainbow, SPHINCS+, RSA, and FALCON.

\begin{figure}[!h]
    \centering
    \includegraphics[width=0.48\columnwidth]{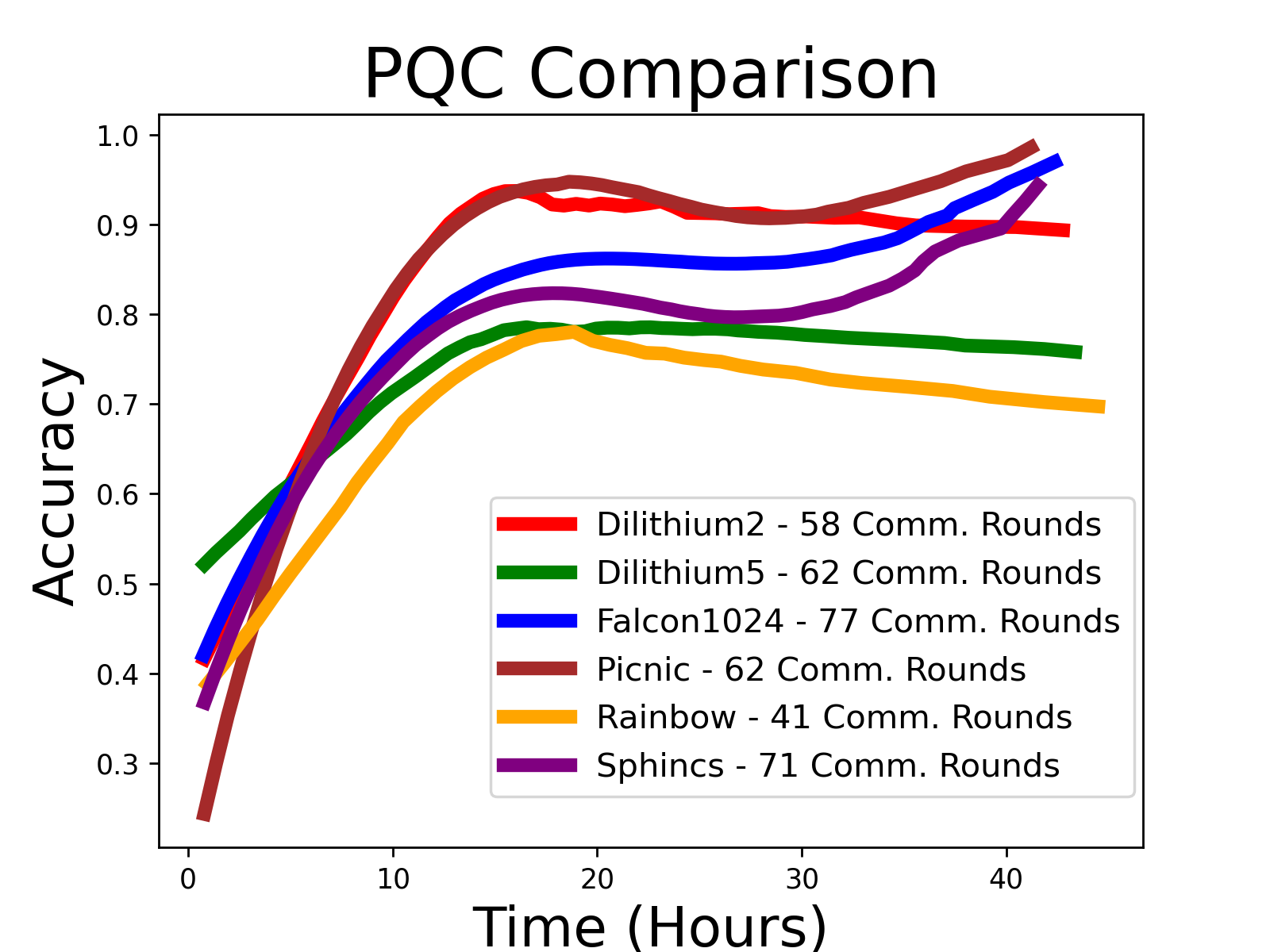}
    \caption{Accuracy comparision of the PQC Schemes}
    \label{fig:pqc-schemes}
\end{figure}

In Figure \ref{fig:pqc-schemes}, we have findings on communication overhead
and the impact on overall system performance. 
The experiment was carried out for around 40 hours.  In that duration, we can see that with each PQC implementation, the number of communication rounds that are completed is different with different schemes. 
For example, Falcon was able to complete 77 communication rounds, while Rainbow only completed 41 communication rounds. 
The difference is similar to that of other PQC schemes.
This clearly implies that, indeed, new PQC schemes work differently, and thus their implementation requires a thorough examination of why they behave as such.
In terms of performance, the more iterations or communication rounds, the better the performance, faster and better.
This is shown by the Falcon and Picnic and SPHINCS+ schemes.
However, the Dilithium5 and Rainbow schemes seem to have suffered in providing better overall system performance.
Thus, we can conclude the necessity of intensive research work towards PQC implementation in modern systems to understand their behavior and
work towards achieving crypto agility.

\section{Conclusion} \label{sec-conclusions}
We proposed a hybrid signature approach for the post-quantum security of BFL.  For improving performance, we developed a multi-factor dependent fuzzy logic for the role selection mechanism in a multi-role environment of BFL.
We presented the idea of using VRF to assist in the consensus mechanism and developed new ways to implement PQC over BFL. We have extensively evaluated the proposed PQC approach in several BFL environments and with different scenarios. Our experimental simulation indicates the importance and practicality of the proposed approaches.